\documentclass[numberedappendix]{emulateapj}
\usepackage{apjfonts}
\usepackage{epsf}

\slugcomment{Accepted in ApJ}

\newcommand\Rfold{R_{\rm fold}}
\newcommand\Rcusp{R_{\rm cusp}}
\newcommand\Afold{A_{\rm fold}}
\newcommand\Acusp{A_{\rm cusp}}

\newcommand\vect[1]{{\vec #1}}
\newcommand\bx   {\vect{x}}
\newcommand\bby  {\vect{y}}
\newcommand\bme  {\vect{\eta}}
\newcommand\bmo  {{\bf 0}}
\newcommand\bt   {\vect{\theta}}
\newcommand\bu   {\vect{u}}
\newcommand\grad {\mathop{\rm grad}\nolimits}
\newcommand\jac  {\mathop{\rm Jac}\nolimits}
\newcommand\ha   {\hat{a}}
\newcommand\hb   {\hat{b}}
\newcommand\hc   {\hat{c}}
\newcommand\hd   {\hat{d}}
\newcommand\Rein {R_{\rm ein}}
\newcommand\order[2]{\mathcal{O}\left(#1\right)^{#2}}

\newcommand\eq[1]{eq.~(\ref{eq:#1})}
\newcommand\refsec[1]{\S\ref{sec:#1}}
\newcommand\refsecs[2]{\S\S\ref{sec:#1} and \ref{sec:#2}}

\begin{document}

\title{Identifying Lenses with Small-Scale Structure.  II.  Fold Lenses}

\author{
  Charles R.\ Keeton\altaffilmark{1},
  B.\ Scott Gaudi\altaffilmark{2},
  and A.\ O.\ Petters\altaffilmark{3,4}
}

\altaffiltext{1}{
  Department of Physics \& Astronomy, Rutgers University,
  136 Frelinghuysen Road, Piscataway, NJ 08837;
  keeton@physics.rutgers.edu
}
\altaffiltext{2}{
  Harvard-Smithsonian Center for Astrophysics, 60 Garden Street,
  Cambridge, MA 02138;
  sgaudi@cfa.harvard.edu
}
\altaffiltext{3}{
  Departments of Mathematics and Physics, Duke University, Durham,
  NC 27708;
  petters@math.duke.edu
}
\altaffiltext{4}{Bass Fellow}

\begin{abstract}

When the source in a four-image gravitational lens system lies
sufficiently close to a ``fold'' caustic, two of the lensed images
lie very close together.  If the lens potential is smooth on the
scale of the separation between the two close images, the
difference between their fluxes should approximately vanish,
$\Rfold \equiv (F_{+}-F_{-})/(F_{+}+F_{-}) \approx 0$.  (The
subscript indicates the image parity.)  Violations of this
``fold relation'' in observed lenses are thought to indicate the
presence of structure on scales smaller than the separation between
the close images.  We present a detailed study of the fold relation
in realistic smooth lenses, finding it to be more subtle and rich
than was previously realized.  The degree to which $\Rfold$ can
differ from zero for smooth lenses depends not only on the distance
of the source from the caustic, but also on its location {\em along}
the caustic, and then on the angular structure of the lens potential
(ellipticity, multipole modes, and external shear).  Since the source
position is unobservable, it is impossible to say from $\Rfold$ alone
whether the flux ratios in an observed lens are anomalous or not.
Instead, we must consider the full distribution of $\Rfold$ values
that can be obtained from smooth lens potentials that reproduce the
separation $d_1$ between the two close images and the distance $d_2$
to the next nearest image.  (By reducing the image configuration to
these two numbers, we limit our model dependence and obtain a generic
analysis.)  We show that the generic features of this distribution
can be understood, which means that the fold relation provides a
robust probe of small-scale structure in lens galaxies.  We then
compute the full distribution using Monte Carlo simulations of
realistic smooth lenses.  Comparing these predictions with the
data, we find that five of the the 12 known lenses with fold
configurations have flux ratio anomalies: B0712+472, SDSS 0924+0219,
PG 1115+080, B1555+375, and B1933+503.  Combining this with our
previous analysis revealing anomalies in three of the four known
lenses with cusp configurations, we conclude that {\em at least half
(8/16) of all four-image lenses that admit generic, local analyses
exhibit flux ratio anomalies}.  The fold and cusp relations do not
reveal the nature of the implied small-scale structure, but do
provide the formal foundation for substructure studies, and also
indicate which lenses deserve further study.  Although our focus is
on close pairs of images, we show that the fold relation can be used
--- with great care --- to analyze all image pairs in all 22 known
four-image lenses and reveal lenses with some sort of interesting
structure.

\end{abstract}


\section{Introduction}
\label{sec:intro}

Once baffling, the flux ratios between the images in four-image
gravitational lens systems have recently become a source of
considerable excitement.  During the 1990s, standard smooth lens
models (using ellipsoidal lens galaxies, plus tidal shear from
lens environments) successfully handled ever-improving data on the
number and relative positions of lensed images, but consistently
failed to fit the image fluxes.  The first step toward solving
this problem came when \citet{MS} realized that small-scale
structure in lens galaxies, which had previously been neglected,
could easily explain the ``anomalous'' flux ratios.
The excitement began in earnest when \citet{MM} and \citet{chiba}
pointed out that the Cold Dark Matter (CDM) paradigm might
naturally explain the sort of substructure required to fit the
fluxes.  Soon after, \citet{DK} introduced a method of analyzing
lens data to measure the properties of substructure.  They concluded
that $2.0_{-1.4}^{+5.0}$ percent (at 90\% confidence) of the mass
in lens galaxies is contained in substructure, which seemed to
agree with CDM predictions, and to reveal that the so-called
``missing'' satellites \citep{bmoore,klypin} are in fact present
but dark.  Anomalous flux ratios had become a powerful test of
CDM on small scales, and potentially a unique probe of the
fundamental nature of dark matter.

Before carrying the conclusions too far, though, we must recall
that there are many links in the chain of logic from observations
of flux ratio anomalies to tests of CDM that need to be filled in.
First, we must identify lenses with anomalous flux ratios.  Second,
we should list all the different types of small-scale
structure\footnote{The term ``substructure'' seems to have come to
represent the sort of small-scale structure predicted by CDM.  The
term ``small-scale structure'' encompasses more general sorts of
structure such as multipole modes, isophote twists, tidal streams,
etc., so it is our term of choice.} that might create flux ratio
anomalies, and understand what observations or analyses could
distinguish between them.  Third, we must see if present data do
distinguish different types of small-scale structure.  If so, we
can then quantify the amount of small-scale structure present in
real lens galaxies.  By comparing the inferred nature and abundance
of small-scale structure to theoretical predictions, we can test the
CDM paradigm.  Finally, if we can understand how the predictions
depend on the assumption that dark matter is cold and collisionless,
we may be able to use lensing to probe the fundamental properties of
the dark matter particle.

\citet{DK} were the first to construct a realization of the full
chain of logic.  Briefly, they identified anomalous flux ratios as
those that could not be fit with standard lens models.  They focused
on radio flux ratios in order to ignore microstructure associated with
individual stars in lens galaxies, and assumed that the only important
small-scale structure is dark matter clumps of the sort predicted by
CDM.  (They argued that other sorts of small-scale structure, such
as globular clusters and dwarf galaxies, are much less abundant than
the inferred number of CDM clumps.)  They assumed that the amount of
CDM substructure is a universal fraction of the total density, used
the lens observations to place constraints on that fraction, and then
compared their results with predictions from CDM simulations.  Making
the various assumptions was necessary to build the first connection
between lens flux ratios and the nature of dark matter.  However,
questions have been raised about some of them, which prompt us to go
back and reassess each link in the chain.  This evaluation is essential
if we want to claim lensing as a reliable probe of small-scale
structure in the universe.  Moreover, it is intrinsically interesting
because it will lead us to a deeper understanding of diverse topics
in both lensing and structure formation theory.

Let us first consider the CDM end of the chain.  There has been a
surge of interest in refining predictions about substructure.  It
now appears that the substructure mass fraction need not be
universal, but may vary both within a given halo and from one halo
to another \citep[e.g.,][]{chen,zentner,MaoJing,oguri}.  Tidal
forces might be able to destroy dark matter clumps at the small
radii where lensed images typically appear, in which case CDM
might predict {\em too little} substructure to explain observed
flux ratio anomalies \citep{MaoJing,amara}.  If so, we should
consider whether small halos projected along the line of sight
can provide sufficient small-scale structure.  The situation is
unclear, as \citet{chen} claim that the millilensing optical
depth from the line of sight is fairly small, while
\citet{metcalfLOS1,metcalfLOS2} claims that interloping structures
are sufficient to explain flux ratio anomalies.  Another possibility
is that revised analyses of lens data may lower the required amount
of small-scale structure (see below).  A third possibility, of
course, is that lensing and CDM simply disagree about small-scale
structure.  In any case, the important point is that the CDM
predictions are challenging and still somewhat uncertain, and more
work needs to be done.  Mastering the theory involves both technical
issues (numerical resolution) and physical effects (dynamical
friction, tidal disruption), so it is not only essential for
interpreting the lensing results, but also interesting in its own
right.

Now moving to the lensing side, we must first ask whether flux
ratio anomalies are real.  \citet{EW} recently suggested that at
least some of the ``anomalies'' might just be artifacts of certain
assumptions in standard lens models.  Specifically, instead of
assuming the usual elliptical symmetry, they allowed perturbations
from $m=3$ and $m=4$ multipole modes.  Such modes are not only
observed in the luminosity distributions of real galaxies
\citep{bender,saglia,rest}, but also predicted in the mass
distributions of simulated galaxies \citep{heyl,naab}, so it does
not seem unreasonable to allow them in lens models.  \citeauthor{EW}
found that they could fit two of the three ``anomalous'' lenses
they considered, without substructure.  However, \citet{congdon}
found that multipole models fail to explain the strongest anomalies.
Also, \citet{KD} argued that even low-order multipole modes cannot
explain an important statistical property of the ensemble of flux
ratio anomalies: an asymmetry between images that form at minima
of the time delay surface and those that form at saddle points,
such that anomalous minima are almost always brighter than expected
while anomalous saddles are usually fainter than expected.  To its
credit, the CDM substructure hypothesis, and stellar microlensing,
can both explain such an asymmetry \citep{MM,SW,analytics,bradac2}.
However, it is not yet known whether alternative hypotheses could
explain the asymmetry as well.  In a different response to
\citeauthor{EW}, \citet{yoo} recently showed that in PG 1115+080
the Einstein ring image of the quasar host galaxy rules out the
sorts of multipole modes that would be needed to fit the quasar
flux ratios.  This type of analysis is very promising, but it
demands deep, high-resolution, near-infrared observations combined
with a sophisticated modeling analysis, and it must be applied on
a case-by-case basis.

Clearly, a top priority must be to develop methods to determine
whether flux ratio anomalies are real and indicate small-scale
structure.  One approach is to look for new data that cleanly
reveal small-scale structure.  The most unambiguous situation is
the detection of flux perturbations associated with microlensing
by stars in the lens galaxy.  Detecting time variability in
optical fluxes can prove that microlensing occurs
\citep[e.g.,][]{wozniak,s1104}.  Barring that, the next best thing
is to take optical spectra of lensed images and use similarities
or differences between emission line and continuum flux ratios to
distinguish between microlensing, millilensing (a term sometimes
applied to flux perturbations caused by CDM-type substructure),
and errors in the macromodel \citep{moustakas}.  The required
observations are challenging, but the method does appear to be
successful \citep{wisotzki0435-2,metcalf2237,morgan,wayth,hst0924}.
At this point, it is appropriate to note that the ``more data''
program has made it possible to conclude that, whatever their
lensing interpretation may be, flux ratio anomalies are not
electromagnetic phenomena.  Measurements of flux ratios at
different epochs and wavelengths have shown that differential
extinction and scattering cannot explain the unusual observed
flux ratios \citep[see][and Appendix B]{koopmans,KD,chibaIR}.

An alternate approach is to reanalyze existing data.  Traditionally,
flux ratio anomalies have been identified as those that cannot be
fit with certain smooth lens models \citep[e.g.,][]{DK,MZ,KD}.
That analysis is, of course, susceptible to the criticism of being
model-dependent.  Perhaps even more important, it may be sensitive
to certain global symmetries in the popular lens models that lead
to global relations among the magnifications of the four images
\citep{dalal,wittmao00,hunterevans,evanshunter}.  Failure to fit
observed flux ratios may simply indicate failure of the global
symmetries --- which is very different from saying that there must
be small-scale structure.  To circumvent both of these problems,
we would like to develop an analysis that is both {\em local} in
the sense that it only depends on properties of the lens potential
around and between closely-spaced images, and {\em generic} in
the sense that it does not depend on any specific properties of
the types of models that are used to analyze the data.

Fortunately, lens theory has uncovered precisely what we need:
local and generic relations between the magnifications between
certain images in certain configurations.  Specifically, two
images in a ``fold pair'' (defined in \refsec{configs}) should
have magnifications $\mu_A$ and $\mu_B$ that satisfy the
approximation relation $|\mu_A| - |\mu_B| \approx 0$; while
three images in a ``cusp triplet'' should have
$|\mu_A| - |\mu_B| + |\mu_C| \approx 0$.  This cusp relation
played a central role in the analysis by \citet{MS} that led to
the idea that lens flux ratios may probe small-scale structure.
If we want to use the fold and cusp relations today, however, we
must rigorously understand how ``local'' and ``generic'' they
really are, and whether they can actually be used as the basis
of a realistic but robust method for identifying flux ratio
anomalies that indicate small-scale structure.

The standard fold and cusp relations are derived from low-order
Taylor series expansions in the vicinity of a fold or cusp caustic
\citep{BN,mao92,weiss,SEF,PLW,GPf,GPc}.  Formally, they are only
valid when the source lies asymptotically close to a caustic.  We
have undertaken to reexamine the relations in more realistic
settings, when the source sits a small but finite distance from
a caustic, and the lens potential has a variety of nontrivial
but smooth structures such as different radial profiles,
ellipticities, octopole ($m=4$) modes, and external tidal shears.
In \citet[hereafter Paper I]{cuspreln}, we studied the cusp
relation.  We found it to be insensitive to the radial profile of
the lens galaxy, but quite sensitive to ellipticity, multipole
modes, and shear.  We quantified the degree to which these features
can cause $|\mu_A| - |\mu_B| + |\mu_C|$ to deviate from zero even
for smooth lenses.  We then compared those allowed deviations with
observed data and found that five observed lenses violate the
realistic cusp relation in ways that indicate the presence of
small-scale structure (B0712+472, RX J0911+0551, SDSS 0924+0219,
RX J1131$-$1231, and B2045+265).  We were very careful to state
the limitations of the analysis, in particular to remark that
study of the cusp relation --- like all other analyses of
single-epoch, single-band flux ratios --- cannot reveal the nature
of the implied small-scale structure.  The strongest conclusion
that can be drawn from a generic analysis is that the lens must
have significant structure on scales smaller than the separation
between the triplet of cusp images.  We believe that this sort of
deep discussion of the general features and applicability of a
generic magnification relation is as valuable as the specific
identification of flux ratio anomalies that it allows.

In this paper, we turn our attention to the fold relation.  We
again seek to understand the general properties of the relation in
realistic situations, and to use that understanding to identify
violations of the fold relation.  We adopt the same basic approach
as in Paper I: we first examine simple lens potentials analytically,
then develop a Monte Carlo approach to study the fold relation in a
realistic lens population, and finally use the realistic fold relation
to look for flux ratio anomalies in observed lenses.  However, many
fine points of the discussion are rather different, because there
are subtle but important ways in which the fold and cusp relations
behave differently.  In addition, we have come to understand that
the fold relation may be used --- with great care --- to learn
something interesting about image pairs that are not obviously fold
pairs.  The discussion of observed lenses therefore has a somewhat
larger scope in this paper than it did in Paper I.  One final
difference is that the sample of published four-image lenses has
grown by three since Paper I.

As in Paper I, we assert that, even though we adopt specific families
of lens potentials, our analysis is more general than explicit
modeling.  One reason is that we take pains to understand what is
generic in the fold relation.  A second reason is that we have a
better distinction between global and local properties of the lens
potential.  For example, a global $m=1$ mode (i.e., a lack of
reflection symmetry) would affect conclusions about anomalies in
direct modeling, but not in our analysis.  A third point is that
our results are less {\em modeling} dependent, less subject to the
intricacies of fitting data and using optimization routines.  A
fourth advantage of our analysis is that, rather than simply showing
the standard models fail to fit a lens, it clearly diagnoses why.
We believe that these benefits go a long way toward establishing
that small-scale structure in lens galaxies is real and can be
understood.

We must address a question that is semantic but important:
Where do we draw the line between a normal ``smooth'' lens
potential and ``small-scale structure''?  As in Paper I, we take
a pragmatic approach and consider ``smooth'' lenses to include
anything that is known to be common in galaxies, especially
early-type galaxies: certain radial density profiles, reasonable
ellipticities, small octopole modes representing disky or boxy
isophotes, and moderate tidal shears from lens environments.
Finding evidence for any or all of those in any given lens would
not cause much stir.  We then consider ``small-scale structure''
to be anything whose presence in lens galaxies would be notable
and worthy of further study.  In other words, we do not attempt
to distinguish between microlensing, CDM-type substructure, massive
and offset disk components, large-amplitude or intermediate-scale
multipole modes, etc.\ as explanations for flux ratio anomalies.
That is properly the subject of a separate analysis, which can begin
only once flux ratio anomalies have been rigorously identified.

The organization of the paper is as follows.  In \refsec{configs}
we introduce a way to quantify four-image lens configurations that
is convenient for the fold relation.  In \refsec{relns} we review
the ideal magnification relations for folds and cusps.  In
\refsec{general} we use a simple lens potential to examine the
general properties of the fold relation in different regimes.  In
\refsec{sims} we introduce a Monte Carlo technique for characterizing
the fold relation for a realistic population of lens potentials.
In \refsec{obs} we use our understanding of the fold relation to
evaluate all of the observed four-image lenses.  We offer our
conclusions and discussion in \refsec{concl}.  Two appendices
provide supporting technical material.  In Appendix A we present
an important extension of the usual Taylor series analysis leading
to the fold relation.  In Appendix B we summarize the data that we
analyze for all of the published four-image lenses.

\section{Characterizing Four-Image Lenses}
\label{sec:configs}

At least 23 quadruply imaged lens systems are known.  This count
includes the 10-image system B1933+503, which is complex because
there are two different sources that are quadruply imaged and a
third that is doubly imaged \citep{sykes}.  It excludes
PMN J0134$-$0931 and B1359+154 because each system has multiple
lens galaxies that lead to image multiplicities larger than four
\citep{1359,0134a,0134b}.  The count also excludes systems like
Q0957+561 in which some faint secondary features, including the
host galaxy of the source quasar, may be quadruply imaged but
are difficult to study \citep{0957bern,0957host}.  Published data
for the quadruply imaged systems are reviewed in Appendix B.

The image configurations of quad lenses can usually be classified
``by eye'' into three categories: folds, cusps,\footnote{Some
authors subdivide cusps depending on whether they are associated
with the long or short axis of the lens potential
\citep[e.g.,][]{saha}, and we will follow suit when convenient.}
and crosses.  The names are related to the location of the source
with respect to the lensing caustics.  For our purposes it is more
important to find a simple but quantitative way to characterize
the configurations.  When studying cusp lenses in Paper I, we
used the separation and opening angle of a triplet of images.  To
study fold configurations, we are interested in pairs of images,
in particular pairs comprising one image at a minimum of the time
delay surface and one at a saddle point.  (The parities of the
images can usually be determined unambiguously; see \citealt{saha}.)
Let us label the two minima $M_1$ and $M_2$, and the two saddles
$S_1$ and $S_2$.  (For definiteness, suppose $M_1$ is the brighter
minimum and $M_2$ the fainter, and likewise for the saddles.)
When considering the pair $M_1 S_1$, for example, we define
$d_1 = D(M_1,S_1)$ and $d_2 = \min[ D(M_1,S_2), D(M_2,S_1) ]$,
where $D(i,j)$ is the distance between images $i$ and $j$.  In
other words, $d_1$ is the separation between the images for the
pair in question, and $d_2$ is the distance to the next nearest
image.  Note that $d_1$ and $d_2$ describe a {\em pair} of images.
At times it is convenient to characterize the full configuration
of all four images, and we define $d_1^*$ and $d_2^*$ to be the
values of $d_1$ and $d_2$ for the pair with the smallest separation.
In other words, a given four-image lens is fully characterized by
the four values of $(d_1,d_2)$ for the four different minimum/saddle
pairs; but it is sometimes convenient to use $(d_1^*,d_2^*)$ as an
abbreviation that encodes the overall morphology of the lens.

\begin{figure}[t]
\centerline{\epsfxsize=2.8in \epsfbox{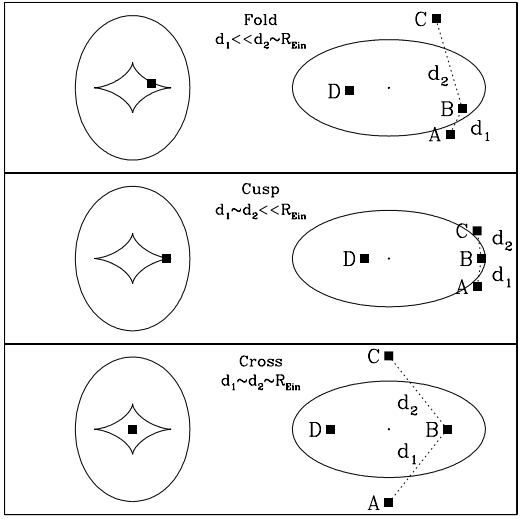}}
\caption{
Three fiducial configurations of four-image lenses: fold {\em (top)},
cusp {\em (middle)}, and cross {\em (bottom)}.  In each panel, the
figure on the left shows the caustics and source position in the
source plane, while the figure on the right shows the critical
curves and image positions in the image plane.  Despite appearances,
the fold and cusp sources sit a finite distance from the caustic.
The configurations are distinguished by the distances $d_1$ and $d_2$,
as indicated.
}\label{fig:f1}
\end{figure}

Figure~1 illustrates the three fiducial configurations, and
indicates $d_1$ and $d_2$ for sample image pairs.  In a fold lens,
the source sits near a fold caustic so two of the images lie close
together with $d_1 \ll d_2$.  Furthermore, $d_2$ is comparable to
the other scale in the problem, the Einstein radius $\Rein$.  In
a cusp lens, the source is near a cusp caustic so three of the
images are close together and we have $d_1 \sim d_2 \ll \Rein$.
If the source does not lie near a caustic, then the images form
a relatively symmetric cross configuration with
$d_1 \sim d_2 \sim \Rein$.

Incidentally, the three ``archetypal'' image configurations shown
in Figure~1 were created using a singular isothermal ellipsoid
lens with axis ratio $q=0.5$ or ellipticity $e=1-q=0.5$.  We chose
source positions such that
$d_1^{\rm fold} = d_1^{\rm cusp} = d_2^{\rm cusp} = 0.46\,\Rein$,
which is similar to the separation between close image pairs and
triplets in observed fold and cusp lenses.  The values
$d_1^{\rm cross} = d_2^{\rm cross} = 1.54\,\Rein$ were set by the
choice of ellipticity.  We then chose the fold source position
such that $d_2^{\rm fold} = d_1^{\rm cross} = d_2^{\rm cross}$.
Having different distances be equal to each other means that we
can smoothly morph from the fold to the cusp by fixing $d_1$ and
varying $d_2$, or from the fold to the cross by fixing $d_2$ and
varying $d_1$.

\section{Asymptotic Magnification Relations for Folds and Cusps}
\label{sec:relns}

In this section we briefly review the expected relations between
the magnifications of images corresponding to a source near a
fold or cusp caustic.  The relations have been discussed before
\citep{BN,mao92,weiss,SEF,PLW,GPf,GPc}, but we have extended the
relations to a higher order of approximation.

As Paper I discussed in depth, when the source lies near a cusp
caustic, the three associated images should have\footnote{In Paper
I we used the absolute value of $\Rcusp$, but it has become clear
that the sign is an important component of theoretical predictions
\citep[e.g.,][]{MM,SW,analytics,KD,bradac2} so we retain it now.
Working with the signed quantity would not change the conclusions
of Paper I.}
\begin{equation} \label{eq:Rcusp}
  \Rcusp
  \equiv \frac{|\mu_A| - |\mu_B| + |\mu_C|}{|\mu_A| + |\mu_B| + |\mu_C|}
  = \frac{F_A - F_B + F_C}{F_A + F_B + F_C}
  \approx 0 ,
\end{equation}
where $\mu_i$ is the signed magnification of image $i$, while
$F_i = F_{\rm src} |\mu_i|$ is the flux of the image if the source
has flux $F_{\rm src}$.  ($\Rcusp$ is defined such that it is
independent of $F_{\rm src}$.)  In our naming convention, B is
the middle of the three images and there is no need to specify
whether it is a minimum or saddle image.  To state \eq{Rcusp}
more precisely, we expand the lens mapping in a Taylor series
about the cusp and find $\Rcusp = 0 + \Acusp\,d^2 + \ldots$, where
$d$ is the maximum separation between the three images, while
$\Acusp$ depends on properties of the lens potential at the cusp
point (physically, what matters is the ellipticity, higher-order
multipoles, and external shear; see Paper I).  Since the constant
and linear terms vanish, a source lying sufficiently close to the
cusp produces three close images with $d \to 0$ and hence
$\Rcusp \to 0$.  As the source moves a small but finite distance
from the cusp, the cusp relation picks up a correction term at
second order in $d$.  Nevertheless, for realistic distributions
of ellipticity, multipole amplitudes, and shear, it is possible
to derive reliable upper bounds on $\Rcusp$.  Roughly speaking,
we may say that those bounds can be violated only if the lens
potential has significant structure on scales smaller than the
distance between images, although Paper I provides a much more
careful discussion.

Appendix A of this paper shows that when the source lies near a
fold caustic, the two images near the fold critical point should
have
\begin{equation} \label{eq:Rfold}
  \Rfold \equiv
  \frac{|\mu_{\rm min}| - |\mu_{\rm sad}|}{|\mu_{\rm min}| + |\mu_{\rm sad}|}
  = \frac{F_{\rm min} - F_{\rm sad}}{F_{\rm min} + F_{\rm sad}}
  \approx 0 .
\end{equation}
We are interested in pairs consisting of a minimum and a saddle,
and we define $\Rfold$ such that the saddle image gets the minus
sign in the numerator.  Again, to be more precise we use a Taylor
series expansion of the lens mapping near the fold point to find
$\Rfold = 0 + A'(\Delta u)^{1/2} + \ldots$, where $\Delta u$ is
the perpendicular distance of the source from the caustic, and
$A'$ is a constant that depends on local properties of the lens
potential (see eq.~\ref{eq:Rfold-src} in Appendix A).  Working
instead in the image plane, we can write
$\Rfold = 0 + \Afold\,d_1 + \ldots$, where $d_1$ is the distance
between the two images (see eq.~\ref{eq:Rfold-img}).  The ideal
fold relation $\Rfold \to 0$ holds only when the source is
asymptotically close to the caustic.  Now there is a correction
term at first order in $d_1$, whose coefficient $\Afold$ depends
on properties of the lens potential (see eq.~\ref{eq:Afold}).  In
other words, the correction to the fold relation is of lower order
than the correction to the cusp relation, which means that the
fold relation is more sensitive to a small offset from the caustic.
Thus, some care will be needed to determine whether an observed
violation of the ideal fold relation really reveals small-scale
structure, or just indicates that the source lies a finite distance
from a fold caustic.

\section{Understanding the Fold Relation}
\label{sec:general}

\subsection{In the asymptotic regime}
\label{sec:gen-asymp}

We can begin to understand general features of the fold relation
by examining the coefficient in the asymptotic limit for $\Rfold$,
\begin{equation}
  \Afold 
  = \frac{3 \psi_{122}^2 - 3 \psi_{112} \psi_{222} + \psi_{2222} (1-\psi_{11})}
    {6 \psi_{222} (1-\psi_{11})}\ ,
\end{equation}
where the $\psi$'s represent various derivatives of the lens
potential, evaluated at the fold point (see
eqs.~\ref{eq:Afold}--\ref{eq:gdef} in Appendix A).  Imagine moving
along the caustic and evaluating $\Afold$ at various points.  As
we approach a cusp, $\psi_{222} \to 0$ while the other derivatives
remain finite \citep[e.g.,][p.~346]{PLW}, so $|\Afold| \to \infty$.
The sign depends on the type of cusp.  A ``positive'' cusp has two
minimum images and one saddle, and typically occurs on the long axis
of the lens potential; it has $\Afold \to -\infty$ with a minus sign
because the saddle image is brighter than each minimum.  A ``negative''
cusp has two saddles and one minimum, and typically occurs on the
short axis of the potential; it has $\Afold \to +\infty$ with a plus
sign because the minimum image is brighter than each saddle.  One
implication of $|\Afold| \to \infty$ is that the fold relation
breaks down near a cusp, but that is not surprising because the
asymptotic analysis in Appendix A explicitly assumes that we have
chosen a fold point and are examining a small neighborhood that
does not include a cusp point.  Besides, near a cusp it is the
cusp relation that ought to be satisfied, not the fold relation.

\begin{figure}[t]
\centerline{\epsfxsize=2.8in \epsfbox{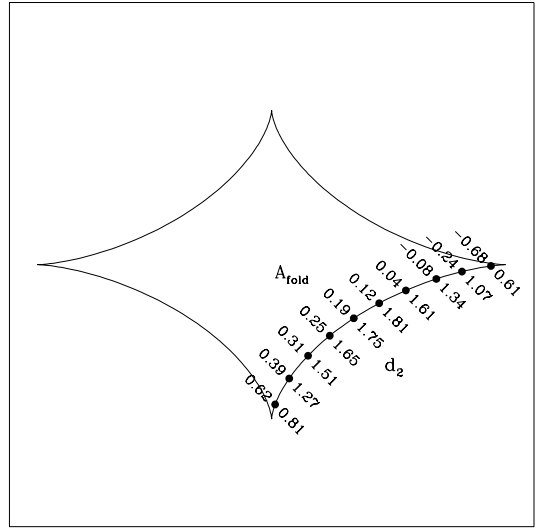}}
\caption{
Caustic curve for an isothermal ellipsoid lens with an axis ratio
$q = 0.5$, or ellipticity $e = 1-q = 0.5$.  The numbers above the
points indicate values of the coefficient $\Afold$ in the asymptotic
fold relation $\Rfold = \Afold\,d_1 + \ldots$ at various points
along the caustic.  (See eq.~\ref{eq:Afold} in Appendix A; recall
that $\Afold$ is to be evaluated {\em on} the caustic, but it then
describes the fold relation in the {\em vicinity} of the caustic.)
The numbers below the points indicate the corresponding values of
$d_2$ (in units with $\Rein=1$).  The other quadrants can be
filled in by symmetry.
}\label{fig:f2}
\end{figure}

The more interesting implication is that $\Afold$ can take on
all real values, both positive and negative.  Unless there is
some remarkable discontinuity, there must be a region where
$\Afold$ changes sign.  Figure~2 confirms that this is the case
for a typical example, namely an isothermal ellipsoid lens with
axis ratio $q=0.5$ or ellipticity $e=1-q=0.5$.  There is a region
where $|\Afold|$ is small or even zero, so that the ideal fold
relation $\Rfold \to 0$ is quite a good approximation.  In this
region the distance $d_2$ is large,\footnote{The distance $d_1$
can be arbitrarily small depending on how close the source is
placed to the caustic, but $d_2$ remains finite even when the
source lies right on the caustic.} but interestingly the smallest
values of $|\Afold|$ do not correspond to the largest values of
$d_2$.  Over the larger range where $d_2$ is large enough that the
image configuration would be classified as a fold (roughly
$d_2 \gtrsim 1$), we find $|\Afold| \sim 0.1$--0.3.  The important
implication is that a lot of lenses that are clearly folds may
nevertheless fail to satisfy the ideal fold relation $\Rfold \to 0$.

An even more important conclusion is that the validity of the
ideal fold relation depends not just on whether the source is
close to a fold caustic, but where the source is located
{\em along} the caustic.  This point is shown more directly in
the next subsection.

\subsection{Across the source plane}
\label{sec:gen-exact}

\begin{figure*}[t]
\centerline{\epsfxsize=6.0in \epsfbox{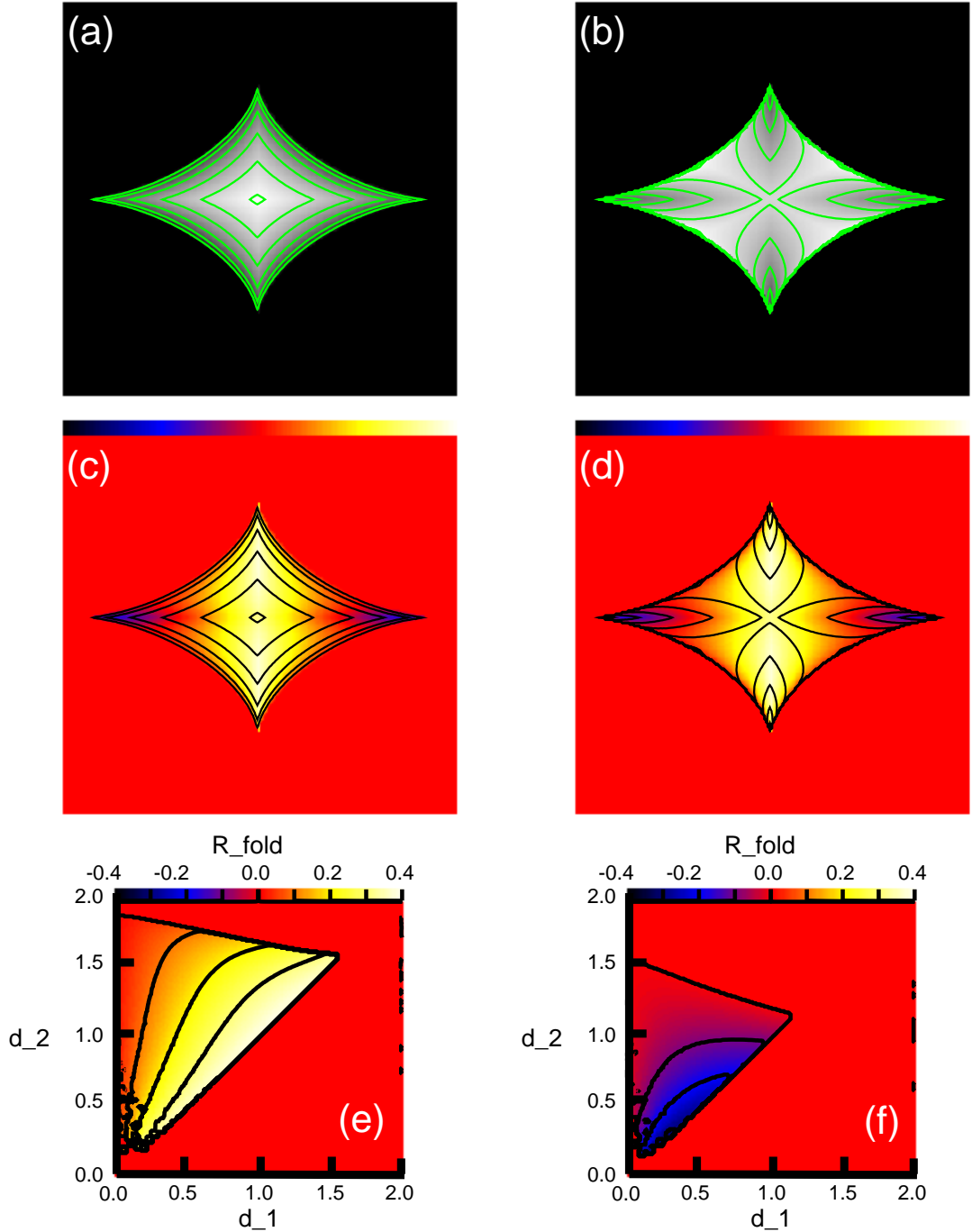}}
\caption{
{\em (a)}
The grayscale and contours both show the distance $d_1$ as a
function of source position, for an isothermal ellipsoid lens
with ellipticity $e=0.5$.  The contours range from 0 to 1.5 in
steps of 0.25, in units with $\Rein=1$.
{\em (b)}
The distance $d_2$ as a function of source position; the contours
are again spaced by 0.25.
{\em (c)}
The colors show $\Rfold$ as a function of source position; the
color coding is shown along the top edge of panel e.  The $d_1$
contours from panel a are overlaid.
{\em (d)}
The colors again show $\Rfold$, with the $d_2$ contours from panel
b now overlaid.
{\em (e--f)}
The colors and contours both show $\Rfold$ as a function of $d_1$
on the abscissa and $d_2$ on the ordinate; the contours range from
$-0.4$ to 0.4 in steps of 0.1.  Panel e shows the case $\Rfold>0$,
while panel f shows the case $\Rfold<0$.  Recall that folds have
$d_1 \ll d_2 \sim \Rein$, cusps have $d_1 \sim d_2 \ll \Rein$, and
crosses have $d_1 \sim d_2 \sim \Rein$.  The region outside the
triangles is inaccessible for this lens potential.  Note that the
figures were generated with Monte Carlo simulations, and the sampling
is imperfect at $d_1 \sim d_2 \ll \Rein$.
}\label{fig:f3}
\end{figure*}

To move beyond the asymptotic regime, we use the software by
\citet{lenscode} to solve the lens equation exactly throughout
source plane for an isothermal ellipsoid lens with ellipticity
$e=0.5$.  For each source inside the astroid caustic, we find
the four images, identify the pair with the smallest separation,
and then compute $d_1$, $d_2$, and $\Rfold$ for that pair.  (These
are by definition the same as $d_1^*$ and $d_2^*$.)  The results
are shown in Figure~3.  First, it is valuable to understand how
$d_1$ and $d_2$ vary with source position, as shown in panels (a)
and (b).  The separation $d_1$ between the images measures very
directly the distance of the source from the caustic.  The distance
$d_2$ to the next nearest image varies along the caustic, and
basically measures the distance of the source from a cusp.  In
general, fixing both $d_1$ and $d_2$ fixes the source to one of
eight positions (two in each quadrant).

Panels (c) and (d) show $\Rfold$ as a function of source position,
with contours of $d_1$ and $d_2$ overlaid.  $\Rfold$ is large and
negative near the long-axis cusp, and it is large and positive in
a band extending from one short-axis cusp to the other and passing
through the origin.  The area in which $\Rfold>0$ is larger than
the area in which $\Rfold<0$, which means that the distribution of
$\Rfold$ values is not symmetric about $\Rfold=0$.  Near the origin,
$\Rfold$ is large and positive.  Both of these points will be
important for our analysis of real lenses in \refsec{obs}.  There
is a ``wedge'' of small $\Rfold$ values starting at the caustic but
extending well inside; this corresponds to the region where the
asymptotic coefficient $\Afold$ is nearly zero (see Fig.~2).
Interestingly, in the region near the caustic and mid-way between
the cusps, where $d_2$ is large and where we would expect to find
archetypal folds, $\Rfold$ is not terribly small.  The source must
get very close to the caustic before $\Rfold$ vanishes.  A remarkable
visual impression is that $\Rfold$ seems to be more correlated with
the $d_2$ contours than with the $d_1$ contours.

Finally, by tabulating the results for all the different source
positions we can plot $\Rfold$ in the $(d_1,d_2)$ plane, as shown
in panels (e) and (f).\footnote{We generated the figures with
Monte Carlo sampling of the source plane, which yields imperfect
sampling of the $(d_1,d_2)$ plane in the lower left corner.}
There are two plots because there are two source positions in each
quadrant, and hence two values of $\Rfold$, with the same values
of $d_1$ and $d_2$.  These figures show more clearly that $\Rfold$
vanishes as $d_1 \to 0$, but the speed with which that occurs
depends on the value of $d_2$.  Furthermore, in the upper left
corner (the region of fold configurations), the $\Rfold$ contours
bend over and become quite sensitive to $d_2$.

We are forced to conclude that the fold relation depends not only
on proximity to a fold caustic, but also on location along the
caustic.  Although we are not shocked --- we knew that the fold
relation should break down near a cusp --- we are nevertheless
surprised to discover how sensitive $\Rfold$ is to location
along the fold caustic even when the source is far from a cusp.
This point is profound, because the location of the source along
the caustic is not observable, and cannot really be determined
from the properties of the two images in the fold pair; it can
only be inferred by considering the properties of the other two
images as well.  In particular, the distance $d_2$ to the next
nearest image gives some indication of the location of the source
along the caustic, and therefore plays a strong role in the fold
relation.

We begin to suspect that using the fold relation in practice is
not a simple matter of finding a close pair of images and asking
how much they deviate from $\Rfold \approx 0$; the fold relation
is in truth more subtle and rich.

\subsection{For all four image pairs}
\label{sec:gen-all}

So far, among the four images in a given configuration we have
only examined the pair with the smallest separation, because the
fold relation best describes close pairs.  The formalism can be
applied to any pair, however, and to round out our general
understanding of the fold relation it is instructive to examine
all the pairs.

\begin{figure}[t]
\centerline{\epsfxsize=3.1in \epsfbox{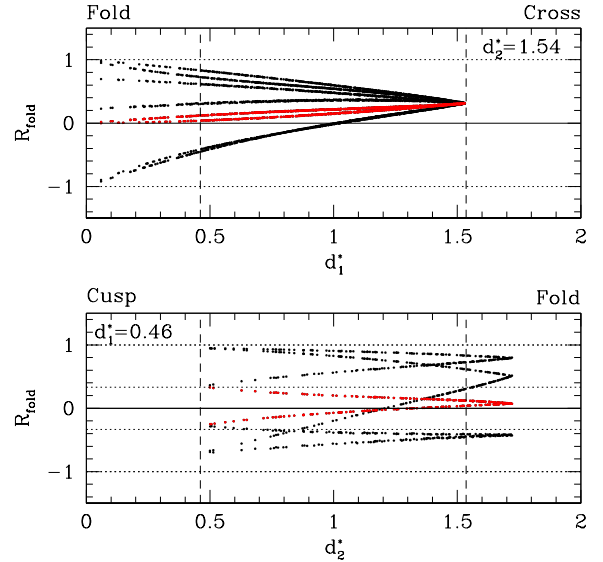}}
\caption{
{\em (Top)}
$\Rfold$ as a function of the distance $d_1^*$, for fixed
$d_2^* = 1.54$.  Moving from left to right smoothly changes the
image configuration from fold to cross.  There are multiple curves
because there are four image pairs for each image configuration,
and there may be two different configurations with the same
$(d_1^*,d_2^*)$.  For each configuration, the smallest-separation
pair is marked in red.
{\em (Bottom)}
$\Rfold$ as a function of the distance $d_2^*$, for fixed
$d_1^* = 0.46$.  Moving from left to right smoothly changes the
image configuration from cusp to fold.
The horizontal lines show various asymptotic limits:
$\Rfold \to 0$ for an ideal fold pair;
$\Rfold \to \pm1/3$ for the two pairs of an ideal cusp triplet; and
$\Rfold \to \pm1$ for two other pairs in an ideal fold or cusp lens.
The vertical lines indicate the separations for our archetypal
lenses:
$d_1^{\rm *fold} = d_1^{\rm *cusp} = d_2^{\rm *cusp} = 0.46$, and
$d_2^{\rm *cusp} = d_1^{\rm *cross} = d_2^{\rm *cross} = 1.54$.
We use an isothermal ellipsoid lens with ellipticity $e=0.5$, and
quote all lengths in units of $\Rein$.
}\label{fig:f4}
\end{figure}

Figure~4 shows $\Rfold$ for all minimum/saddle image pairs, as a
function of the distances $d_1^*$ and $d_2^*$ that characterize
the image configuration.  In the top panel we fix $d_2^*$, so
varying $d_1^*$ morphs the configurations from folds to crosses.
The largest value of $d_1^*$ corresponds to a symmetric cross, in
which case the two minima are identical and the two saddles are
identical, so all four minimum/saddle pairs have the same value
of $\Rfold$.  In the limit $d_1^* \to 0$ we obtain ideal folds,
and the fold pair converges to $\Rfold \to 0$ (the ideal fold
relation).  In this limit two other pairs converge to
$\Rfold \to \pm1$, which is easily understood: the two fold
images (A and B in Fig.~1) have much higher magnifications than
the two other images, so the pairs AD and CB will both have
$\Rfold \to \pm1$.  There is no intuitively obvious asymptotic
limit for the pair of non-fold images (CD).  The figure suggests
that such limits do exist, but we suspect that they depend on
properties of the lens potential in ways that the limits
$\Rfold \to 0$ and $\pm1$ do not.

In the bottom panel we fix $d_1^*$, so varying $d_2^*$ morphs
the configurations between folds and cusps.  Here the fold limit
does not quite reach $\Rfold \to 0$ and $\pm1$ because we have
fixed $d_1^*$ to a finite value that does not actually correspond
to an ideal fold.  The more interesting limits are in the
direction of cusps.  As $d_2^* \to d_1^*$ we obtain a symmetric
cusp configuration.  For a symmetric ideal cusp, we can predict
$\Rfold \to \pm1/3$ and $\pm1$ based on the following logic.
By symmetry, $F_A = F_C$ so the ideal cusp relation implies
$F_B \approx 2 F_A$, and the fold relation then yields
$\Rfold \approx \pm1/3$.  The sign is $+$ for a long-axis cusp
(in which case B is a saddle), or $-$ for a short-axis cusp
(B is a minimum).  At the same time, in an ideal cusp the images
A, B, and C are all much brighter than the fourth image D, so any
pair involving D has $\Rfold \approx \pm1$.  As the source moves
around the numerical values will change, but we generically expect
two distinct values of $\Rfold$, one positive and one negative,
for cusp lenses.  Our archetypal cusp lens does not quite reach
the asymptotic values because the source lies a finite distance
from the cusp ($d_1^*$ is 0.46 rather than $\approx 0$), but it
does confirm the basic reasoning.

Examining $\Rfold$ for widely-separated image pairs in this way
does not really tell us about small-scale structure, because we
are no longer restricted to short length scales.  Nevertheless,
it is still helpful for obtaining a general understanding of the
fold relation.

\section{The Fold Relation in Realistic Lens Potentials}
\label{sec:sims}

While the ideal fold relation $\Rfold \to 0$ is completely general,
it is only valid when the source is extremely close to a caustic.
In realistic situations, the better approximation
$\Rfold = \Afold\,d_1$ depends on the source position and
properties of the lens potential.  In Paper I, we explicitly
showed that the properties of the lens potential affecting the
cusp relation are ellipticity, low-order multipole modes, and
tidal shear.  Here, we simply {\em define} a ``realistic smooth
lens'' to be one that has these angular structures.  (See
\refsec{intro} for more discussion.)  Unfortunately, the
ellipticity, multipole moments, and shear in individual lenses
cannot be observed directly.  Ellipticity and multipole modes
in the lens galaxy {\em light} may be measurable, but for lensing
we need the properties of the {\em mass}.  The mass properties
could be constrained with lens modeling \citep[with perhaps the
best example being the analysis by][]{yoo}, but we seek to avoid
model dependence as much as possible.  Instead, our approach is
to adopt observationally motivated priors on the distribution
of ellipticity, multipole modes, and shear and use Monte Carlo
simulations to derive probability distributions for $\Rfold$
for a realistic lens population.

\subsection{Methods}
\label{sec:sims-meth}

The simulation methods are the same as in Paper I, so we review
the main points here and refer the reader to that paper for
further details.  We consider only isothermal radial profiles
($\Sigma \propto R^{-1}$) for the simulated galaxies, because in
Paper I we showed that local analyses of the lens mapping are not
very sensitive to changes in the radial profile.  For the angular
structure, we consider ellipticity as well as additional octopole
modes ($m=4$ multipole perturbations).  To model populations of
early-type galaxies, we draw the ellipticities and octopole
moments from measurements of isophote shapes in observed samples
of early-type galaxies.\footnote{Most lenses are produced by
early-type galaxies.  Among the four-image lenses, the only known
spiral lens galaxy is in Q2237+0305, and even there the images
are most affected by the spheroidal bulge.  This lens is not very
important for our analysis, because it is a cross lens, and because
it is already known to exhibit microlensing \citep[e.g.,][]{wozniak}.}
Even if the shapes of the light and mass distributions are not
identical on a case-by-case basis, it seems reasonable to think
that their distributions may be similar \citep[see][]{RT}.  Indeed,
the distribution of isodensity contour shapes in simulated merger
remnants is very similar to the observed distribution of isophote
shapes \citep{heyl,naab}.  We use three different observational
samples, because they have different strengths and weaknesses and
allow a check for systematic effects:
\begin{itemize}

\item \citet*{JFK} report ellipticities for 379 E/S0 galaxies in
11 clusters, including Coma.  Their ellipticity distribution has
mean ${\bar e} = 0.31$ and dispersion $\sigma_e = 0.18$.  They do
not report octopole moments.

\item \citet*{bender} report ellipticities and octopole moments
for 87 nearby, bright elliptical galaxies.  Their ellipticity
distribution has mean ${\bar e} = 0.28$ and dispersion
$\sigma_e = 0.15$, while their octopole moment distribution has
mean ${\bar a}_4 = 0.003$ and dispersion $\sigma_{a_4} = 0.011$.

\item \citet*{saglia} report ellipticities and octopole moments
for 54 ellipticals in Coma.  Their ellipticity distribution has
${\bar e} = 0.30$ and $\sigma_e = 0.16$, while their octopole
moment distribution has ${\bar a}_4 = 0.014$ and
$\sigma_{a_4} = 0.015$.

\end{itemize}
The ellipticity and octopole distributions for the three samples
are shown in Fig.~6 of Paper I.  All three samples are limited to
low-redshift galaxies (by the need for good resolution to measure
isophote shapes).  We must assume that the distributions are
reasonable for intermediate-redshift galaxies as well, which seems
plausible if major mergers involving ellipticals are infrequent.

For the external shear amplitude, we adopt a lognormal distribution
with median $\gamma = 0.05$ and dispersion $\sigma_\gamma = 0.2$ dex.
This is consistent with the distribution of shears expected from
the environments of early-type galaxies, as estimated from $N$-body
and semianalytic simulations of galaxy formation by \citet{holder}.
It is broadly consistent with the distribution of shears required
to fit observed lenses.  \citet{DW} use a halo model calculation to
suggest that the median shear should be more like $\gamma = 0.03$.
However, the smaller median shear is not very compatible with the
shears required to fit observed four-image lenses \citep[e.g.,][]{KKS}.
Furthermore, if we want to determine how much $\Rfold$ can deviate
from zero for smooth lens potentials, then the conservative approach
is to adopt the larger median shear.  We assume random shear
directions.

For each combination of ellipticity, octopole moment, and
shear\footnote{Note that we need not specify the galaxy mass,
because for an isothermal lens the mass merely sets the length
scale $\Rein$, and we can always work in units such that $\Rein=1$.}
we choose random sources with density $\sim\!10^{3}\,\Rein^{-2}$,
solve the lens equation using the software by \citet{lenscode},
and compute $(d_1,d_2,\Rfold)$ for each minimum/saddle image pair.
For each input distribution, we examine a total of $\sim\!10^{6}$
mock four-image lenses.  Note that choosing sources with uniform
density in the source plane has two important consequences.  First,
it ensures that each lens potential is automatically weighted by
the correct lensing cross section.  Second, it means that we
neglect magnification bias, which would favor lenses with higher
amplifications, and therefore give more weight to sources near
the caustics that produce {\em small} deviations from the ideal
fold relation.  We therefore believe that neglecting magnification
bias is the conservative approach when seeking to understand how
large the deviations can be for smooth lens potentials.

\subsection{First results}
\label{sec:sims-res}

\begin{figure*}[t]
\centerline{\epsfxsize=6.0in \epsfbox{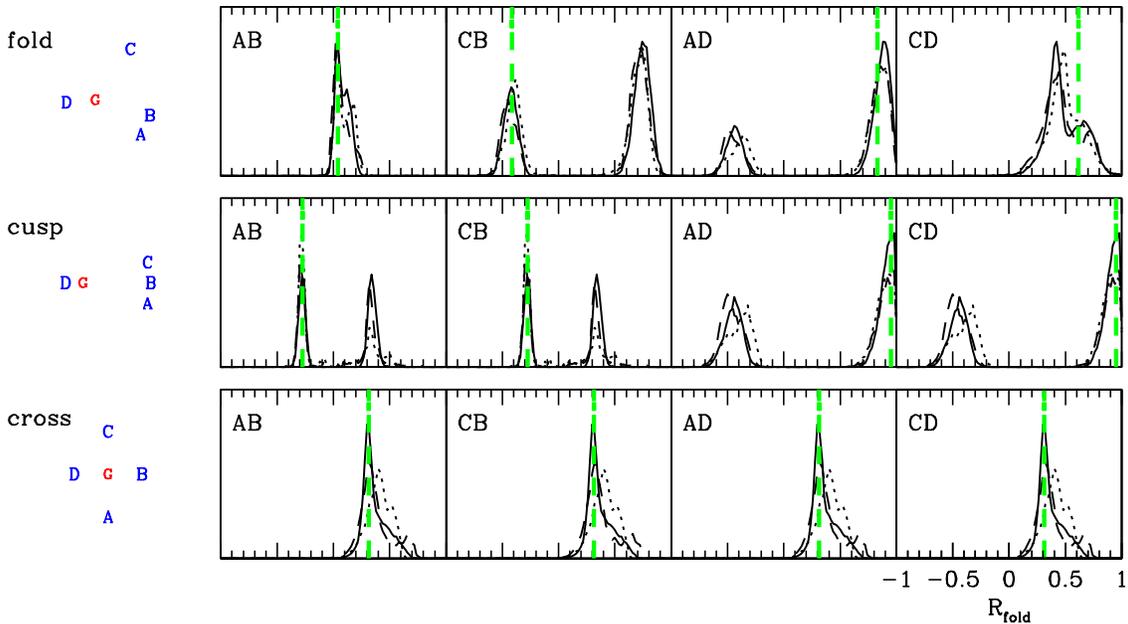}}
\caption{
Probability distributions for $\Rfold$.  On the left we show the
image configurations, with lensed images (A--D) marked in blue and
the lens galaxy (G) marked in red.  The four columns then show
$\Rfold$ for the different image pairs, sorted by increasing $d_1$
from left to right.  (We adopt the convention of naming each pair
such that the first letter indicates the minimum image while the
second letter indicates the saddle.)  The green vertical lines
mark the actual values for our archetypal lenses.  The black
curves show the predicted distributions for realistic lens
populations, with solid, dashed, and dotted curves showing results
for Monte Carlo simulations based on the J{\o}rgensen, Bender, or
Saglia galaxy samples, respectively.  The predicted distributions
are normalized to unit area.
}\label{fig:f5}
\end{figure*}

We use our ensemble of mock image configurations to extract the
probability distribution for $\Rfold$ at fixed values of $d_1$
and $d_2$.\footnote{Strictly speaking, to accommodate our finite
sampling of phase space, we consider all image configurations
within $\pm0.05$ of the specified $d_1$ and $d_2$ values.}
Conceptually, this is like going to the appropriate point in the
$(d_1,d_2)$ plane of Figure~3e-f and reading off $\Rfold$, except
that we now consider a large ensemble of lens potentials.  To
illustrate how we use these distributions, Figure~5 compares the
value of $\Rfold$ for each image pair in our archetypal fold,
cusp, and cross lenses to the appropriate conditional probability
distribution $p(\Rfold | d_1,d_2)$ derived from the Monte Carlo
simulations.  If the observed value lies outside the predicted
distribution, then we conclude that the image pair is inconsistent
with lensing by a realistic population of smooth lens potentials.
It is reassuring to see that our archetypal lenses (which were
generated with a smooth lens) are indeed found to be consistent
with lensing by a smooth potential.

We can observe some of the general features identified in
\refsec{gen-all}.  Many of the distributions are bimodal, and
some of those have two completely disjoint peaks.  This is because
given values of $d_1$ and $d_2$ can correspond to multiple source
positions that yield different $\Rfold$ values (see Fig.~3).
In the fold lens, the pair involving two fold images (AB) has
$\Rfold \approx 0$.  The two pairs involving one fold and one
non-fold image (CB and AD) each have $|\Rfold|$ large; they do
not have $\Rfold \to \pm1$ because the source sits a finite
distance from the caustic, but the general trend that the fold
pair has small $|\Rfold|$ while the two fold/non-fold pairs have
large $|\Rfold|$ is confirmed.  In the cusp lens, two pairs have
$\Rfold \approx \pm1/3$ while the other two have $|\Rfold|$ large.
Again, the reason that the peaks in the AD and CD pairs do not
actually reach $\Rfold \to \pm1$ is because the source sits a
finite distance from the caustic.  Finally, in the cross case
all four pairs have similar $\Rfold$ distributions --- identical
in the case of a symmetric cross --- which are centered at some
positive value but fairly broad.  The consistency between our
general analytic arguments and our detailed Monte Carlo simulations
is reassuring, and indicates that we have obtained new, deep
insights into the fold relation.

\section{Application to Observed Lenses}
\label{sec:obs}

We are finally ready to examine the fold relation for observed
four-image lenses.  We summarize the data here (\refsec{obs-data}),
and provide more details in Appendix B.  Our main interest for the
fold relation is of course fold image pairs (\refsec{obs-fold1}),
but it is also interesting to consider the other image pairs in
fold lenses (\refsec{obs-fold2}), as well as image pairs in cusp
(\refsec{obs-cusp}) and cross (\refsec{obs-cross}) lenses.

\subsection{Summary of the data}
\label{sec:obs-data}

Table~1 lists the values of $d_1$, $d_2$, $\Rein$, and $\Rfold$
for all minimum/saddle image pairs in 22 known four-image lens
systems,\footnote{We do not analyze 0047$-$2808, as discussed
in Appendix B.} and Appendix B provides some comments about the
data.  Most available flux ratio data come from broad-band
optical/near-infrared images or radio continuum observations.  We
consider separate optical and radio $\Rfold$ values, since they
correspond to very different source sizes and therefore provide
different information about small-scale structure.  We also
consider any other flux ratio data that are available: in the
mid-infrared for Q2237+0305 \citep{agol} as well as PG 1115+080
and B1422+231 \citep{chibaIR}, and the optical broad emission
lines for HE 0435$-$1223 \citep{wisotzki0435-2}, WFI 2033$-$4723
\citep{morgan}, and SDSS 0924+0219 \citep{hst0924}.

We need the Einstein radius $\Rein$ to normalize $d_1$ and $d_2$.
This must be determined from lens models, but it is quite robust
and not very dependent on the choice of model
\citep[e.g.,][]{csk91,cohn}.  We treat the main lens galaxy as an
isothermal ellipsoid with surface mass density
\begin{equation}
  \kappa(r,\theta) = \frac{\Sigma(r,\theta)}{\Sigma_{\rm crit}}
    = \frac{\Rein}{2 r [1-\epsilon\cos2(\theta-\theta_\epsilon)]^{1/2}}\ ,
\end{equation}
where $\Sigma_{\rm crit}$ is the critical surface density for
lensing, $\epsilon$ is related to the axis ratio $q$ of the galaxy
by $\epsilon = (1-q^2)/(1+q^2)$, and $\theta_\epsilon$ is the
position angle of the galaxy.  Few four-image lenses can be fit
by a pure isothermal ellipsoid model, because elliptical lens
galaxies tend not to be isolated.  In most cases, modeling the
environmental contribution to the lens potential as an external
shear provides an excellent fit to the image positions
\citep[e.g.,][]{KKS}.  The exceptions are HE 0230$-$2130,
MG J0414+0534, RX J0911+0551, and B1608+656, each of which is
known to have a satellite galaxy near the main lens galaxy that
must be included in order to fit the image positions
\citep{wisotzki0230,SchechterMoore,koopmans1608}.  We treat the
satellite galaxies as isothermal spheres.  We stress that when
fitting the models to determine $\Rein$, we use only the relative
positions of the images and the lens galaxy as constraints; it
is not necessary to use the flux ratios as model constraints.

\begin{figure}[t]
\centerline{\epsfxsize=3.0in \epsfbox{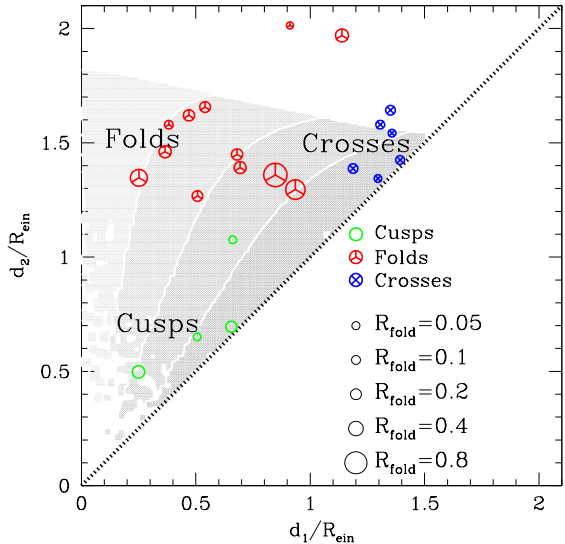}}
\caption{
Colored points mark the locations of known four-image lenses in
the plane of $d_1/\Rein$ and $d_2/\Rein$.  The color indicates the
visual classification as a fold, cusp, or cross, while the point
size indicates the value of $\Rfold$.  To help guide the eye, the
grayscale shows $\Rfold$ for an isothermal ellipsoid lens with
ellipticity $e=0.5$ (from Fig.~3e).
}\label{fig:f6}
\end{figure}

Figure~6 shows how the observed lenses populate the $(d_1,d_2)$
plane.  Although we have labeled them as folds, cusps, and
crosses, in fact there is no sharp distinction between the fold
and cross samples.  SDSS 0924+0219 and B1933+503, which have
$d_1/\Rein \sim 0.9$, could arguably be relabeled as crosses,
although we choose not to do so (see \refsec{obs-fold1}).  The
smooth transition simply reflects the fact that there are no
sharp boundaries between different four-image configurations in
the source plane.

In the figure there is a particular region occupied by simulated
lenses (the grayscale), but it is specific to an isothermal
ellipsoid lens with ellipticity $e=0.5$.  Varying the ellipticity
and/or adding shear would move the upper edge so that the region
could accommodate the other cross lenses.  The two fold lenses
at $d_2/\Rein \sim 2$ are a different story, though.  These are
HE 0230$-$2130 and B1608+656, each of which has two lens galaxies.
Turning this around, we may say that observed lenses that are
outliers in the $(d_1,d_2)$ plane are likely to have complex lens
potentials containing multiple galaxies.

\subsection{Fold image pairs}
\label{sec:obs-fold1}

\begin{figure*}[t]
\centerline{\epsfxsize=6.0in \epsfbox{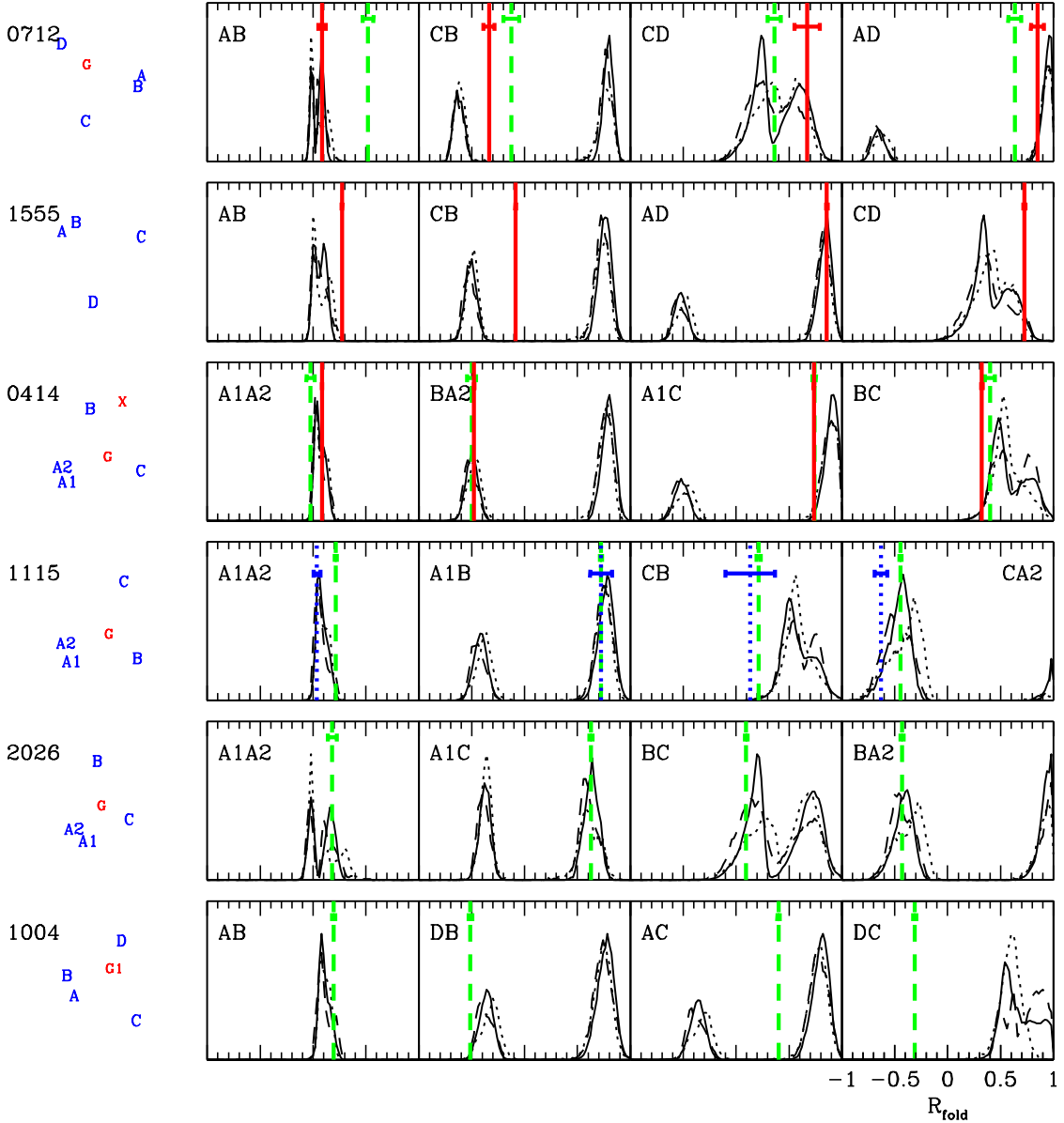}}
\caption{
Observed and predicted $\Rfold$ values, for six of the known fold
lenses.  The black curves show the predicted distributions for
realistic lens populations, as in Fig.~5.  The vertical colored
lines show the observed values and their uncertainties, with green
indicating optical/near-infrared data and red indicating radio
data.  For PG 1115+080, the blue lines indicate mid-infrared data
\citep{chibaIR}.  All data are listed in Table~1.  The lenses are
sorted by increasing $d_1^*/\Rein$ from top to bottom.  For each
lens, the image pairs are sorted by increasing $d_1/\Rein$ from
left to right.
}\label{fig:f7}
\end{figure*}

\begin{figure*}[t]
\centerline{\epsfxsize=6.0in \epsfbox{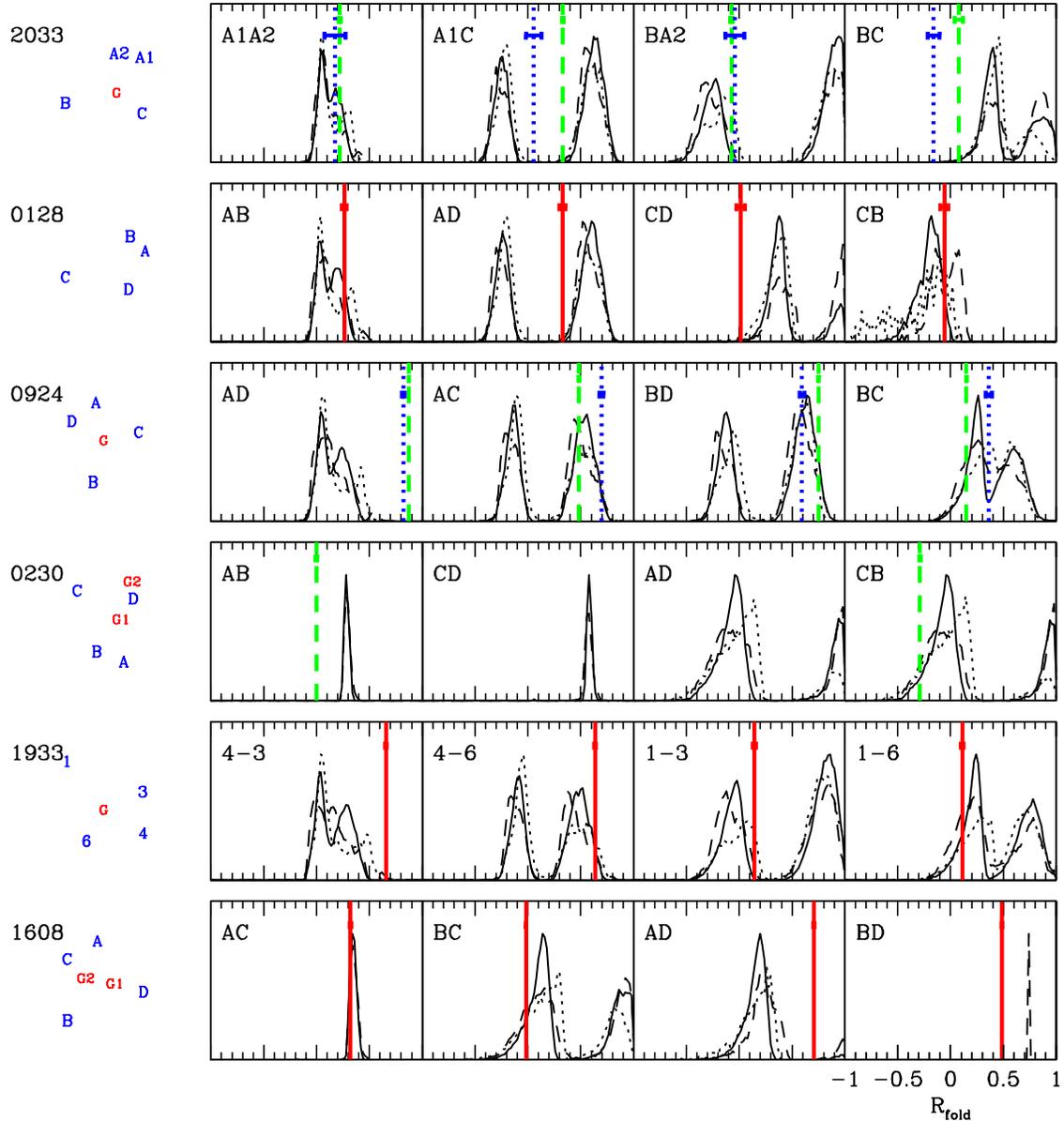}}
\caption{
Similar to Fig.~7, but for the remaining six known fold lenses
(again sorted by $d_1^*/\Rein$).  For SDSS 0924+0219 and
WFI 2033$-$4723, the vertical green lines show data from
broad-band optical flux ratios, while the vertical blue lines
show data from optical emission line flux ratios
\citep{morgan,hst0924}.
}\label{fig:f8}
\end{figure*}

\begin{figure}[t]
\centerline{\epsfxsize=3.1in \epsfbox{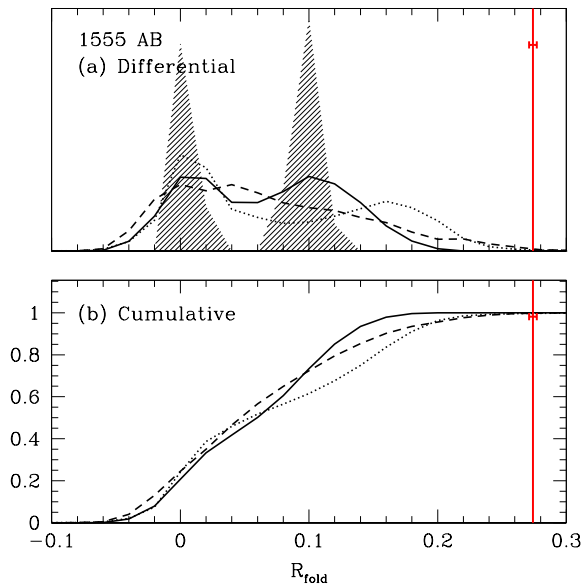}}
\caption{
{\em (a)} Close-up of the AB panel for B1555+375 from Fig.~7.
The red line shows the observed radio value for $\Rfold$, with its
uncertainty (see Table~1).  The solid, dashed, and dotted curves
show the predicted $\Rfold$ distributions for Monte Carlo
simulations based on the J{\o}rgensen, Bender, and Saglia galaxy
samples, respectively.  For comparison, the hatched region shows
the distributions for an isothermal ellipsoid with ellipticity
$e=0.5$.
{\em (b)} Similar to panel a, but showing cumulative probability
distributions.
}\label{fig:f9}
\end{figure}

We now examine the fold relation by comparing the observed
$\Rfold$ values to the distributions expected for a realistic
galaxy population.  Figures~7--8 show the comparisons for the
12 fold lenses,\footnote{Note that we now include B0712+472
among the folds, even though we considered it a cusp in Paper I.
Both classifications seem valid, depending on one's purpose.
The close pair AB can be considered a fold, while the close
triplet ABC can be considered a cusp.  The source must lie
close to the caustic in a region not far from a cusp.  For the
purposes of this paper, it is a fold.} arranged in order of
increasing $d_1^*/\Rein$.  While there is a tremendous amount of
information here, the discussion in \refsec{general} helps us
pick out the main trends.  First let us consider the various
predicted distributions.  When $d_1^*/\Rein$ is small, $\Rfold$
for the fold pair (Column 1) is predicted to lie in a very narrow
range near zero.  This is the fold relation in its familiar form.
At the same time, two other image pairs have distributions that
feature two narrow and well-separated peaks (compare the top row
of Fig.~5), while the fourth pair has a broad distribution with
no particular center.  As $d_1^*/\Rein$ increases, the distribution
for the fold pair broadens while the two peaks for the next closest
pair (Column 2) tend to move closer together.  HE 0230$-$2130 and
B1608+656 buck these trends, for a simple reason: they have two
lens galaxies, so they have configurations that are rare in our
Monte Carlo simulations,\footnote{In fact, the J{\o}rgensen and
Saglia simulations do not contain {\em any} configurations with
the same $d_1$ and $d_2$ values as the BD pair in B1608+656, to
within the sampling resolution of our Monte Carlo simulations.}
and that leads to narrow and unusual predicted $\Rfold$
distributions.

For most image pairs, there are no tremendous differences between
the $\Rfold$ distributions from the three different simulations
(based on the J{\o}rgensen, Bender, or Saglia galaxy samples;
also see Fig.~9).  This gives us confidence that our conclusions
are robust in the sense of not being very sensitive to the
simulation
input data.

Now we turn to the observed values of $\Rfold$.  Many of them
lie within the predicted range, so there is no obvious violation
of the fold relation.  The outliers are as follows:
\begin{itemize}

\item B0712+472:  The optical data grossly violate the fold
relation, but the radio data do not (as in the cusp relation;
see Paper I).  The wavelength dependence suggests that the
optical anomaly is caused by microlensing.

\item B1555+375:  The radio data violate the fold relation at
high confidence, as shown more clearly in Figure~9.

\item PG 1115+080:  The optical value of $\Rfold$ differs from
the predictions at 99.2\% confidence for the J{\o}rgensen and
Saglia simulations, and 96.1\% confidence for the Bender
simulations.  The flux ratios can in principle be fit using
large-amplitude multipole modes \citep{KD}, but such modes are
inconsistent with the Einstein ring image of the quasar host
galaxy \citep{yoo}.  In other words, it appears that this lens
is anomalous, but the model-independent evidence is not quite
as secure as for the other anomalies.  The mid-infrared value
of $\Rfold$ differs from the optical value and agrees well
with the predicted distribution, suggesting that the optical
anomaly is created by microlensing.

\item SDSS 0924+0219:  Although AD is not a particularly close
pair (making the predicted $\Rfold$ distribution fairly broad),
image D is so faint that there is a gross violation of the fold
relation in both broad-band and broad emission line flux ratios.
Differences between the flux ratios plus time variability
suggest the presence of microlensing \citep{csk0924,hst0924}.

\item B1933+503:  Although the 4/3 image pair is not particularly
close, image 4 is so bright that there is a clear violation of
the fold relation in the radio data.

\end{itemize}

HE 0230$-$2130 and B1608+656 deserve mention because each has
two lens galaxies.  In HE 0230$-$2130, the observed $\Rfold$ value
shows images A and B to be more similar than expected for simple
smooth lens potentials.  In B1608+656, the observed $\Rfold$ value
agrees with the predictions for simple lenses, which seems
coincidental.  These systems show that violations of the fold
relation can reveal the lens potential to contain structure that
is complex but not necessarily small-scale.  That raises the
question of whether any of the other anomalies could be caused
by something large like a second galaxy.  Probably not: in both
of these systems the second galaxy was already known from direct
observations and analyses of the image positions (see
\refsec{obs-data}).  In other words, it is difficult for a massive
second galaxy to escape notice.  We therefore believe that the
``second galaxy'' hypothesis is not a valid explanation for most
flux ratio anomalies.

To summarize, our analysis of the fold relation reveals two
flux ratio anomalies that were already known from violations of
the cusp relation (B0712+472 optical, and SDSS 0924+0219 optical).
It also reveals strong new evidence for fold flux ratio anomalies
in B1555+375 (radio) and B1933+503 (radio), plus good but slightly
less strong evidence for a fold anomaly in PG 1115+080 (optical).
In addition, a violation of the fold relation in HE 0230$-$2130
(optical) is presumably due to the presence of a second lens
galaxy in that system.

While these specific conclusions are valuable, there are some
important general lessons as well.  First, even the closest
observed fold image pairs have predicted $\Rfold$ distributions
with a finite width.  Therefore, a non-zero $\Rfold$ value in
the range $\Rfold \sim 0$--0.2 {\em cannot} generally be taken
to indicate a flux ratio anomaly.  As a rule of thumb, when
$d_1/\Rein \lesssim 0.4$ it does appear that a value
$\Rfold \gtrsim 0.2$ is likely to indicate an anomaly, although
we caution that this is just a rule of thumb and a full analysis
of the predicted $\Rfold$ distribution must be done to reliably
identify an anomaly.

The importance of the full analysis becomes clear when we
consider PG 1115+080, WFI 2026$-$4536, and SDSS 1004+4112.  These
three lenses have similar configurations with $d_1/\Rein \approx 0.5$,
and (curiously enough) they all have $\Rfold \approx 0.2$.  Yet one
is anomalous (PG 1115+080), while the other two are fully compatible
with the predicted distributions.  What's more, the predicted
distributions for WFI 2026$-$4536 are bimodal and qualitatively
different from those for the other two lenses, even though all
three image configurations are visually similar.  These lenses
teach the lesson that identifying fold flux ratio anomalies is not
a simple matter of finding a close pair of images and asking whether
$\Rfold \to 0$.  The distance from $\Rfold \to 0$ that is needed
to provide strong evidence for an anomaly depends in a complicated
way on various properties of the lens potential that cannot be
directly observed.  Only a full and careful analysis of the fold
relation can handle these issues.

We conclude that violations of the fold relation can be used to
find flux ratio anomalies in a fairly model-independent way.
However, that analysis is more subtle than was previously
realized.  It is necessary to know not only the separation $d_1$
between the two images, but also the distance $d_2$ to the next
nearest image, and to account for the finite width of the $\Rfold$
distribution expected for smooth lenses.

Finally, it is worthwhile to comment that all twelve of the fold
image pairs have $\Rfold$ values that are positive or consistent
with zero.  An important prediction to emerge from theoretical
studies is that small-scale structure (either dark matter clumps
or stars) should tend to amplify minimum images and/or suppress
saddle images \citep{MM,SW,analytics,bradac2}.  Since either
possibility would make $\Rfold > 0$, seeing only non-negative
values is certainly consistent with the substructure hypothesis
\citep[see][]{KD}.  It is inconsistent with non-gravitational
explanations of flux ratio anomalies (such as extinction or
scattering), because those should affect minimum and saddle images
in the same way.  What is less clear is whether lumpy substructure
is the only thing that can explain the asymmetry between minima
and saddles, or whether small-scale but smooth structure is a
viable alternative.  Our analysis does offer an intriguing hint:
nearly all the weight in our predicted $\Rfold$ distributions lies
at $\Rfold > 0$, which indicates that even smooth, global features
like ellipticity and shear affect minima and saddles differently.
Still, it is not clear whether smooth features can explain the
further asymmetry that all the anomalous $\Rfold$ values {\em exceed}
the predictions.  The minimum/saddle asymmetry appears to be a very
promising probe of small-scale structure, but much more study is
clearly called for.

\subsection{Other image pairs in fold lenses}
\label{sec:obs-fold2}

We can also consider the $\Rfold$ values for the other image
pairs in fold lenses, although we must be careful about how we
interpret them.  As discussed in \refsecs{gen-all}{sims-res},
there are some useful general properties of the fold relation
for these other image pairs.  For example, the predicted $\Rfold$
distribution for a pair comprising a fold image and a non-fold
image has two narrow peaks, one positive and one negative.
While we are not aware of a simple way to predict the specific
values, it seems from Figures~7--8 that they depend mainly on
$d_1^*/\Rein$.

There are several image pairs for which the observed $\Rfold$
value lies far from the peaks in the predicted distributions:
CB in B0712+472, CB in B1555+375, AC and DC in SDSS 1004+4112,
and A$_1$C and BC in WFI 2033$-$4723.\footnote{Note that in
WFI 2033$-$4723 the differences between the optical continuum
flux ratios and the emission line flux ratios are interpreted
as evidence for microlensing \citep[see][]{morgan}.}
The relatively large distance between the images in each pair
prevents us from concluding that the discrepancies reveal
``small-scale'' structure in the lens.  We can still conclude,
though, that each pair is inconsistent with smooth lens models
containing moderate ellipticities, octopole moments, and shears.
Indeed, two of these lenses are already known to have complex
potentials.  SDSS 1004+4112 is produced by a cluster of galaxies
\citep{oguri1004}, while WFI 2033$-$4723 appears to lie in a
group of galaxies with at least six perturbers lying within
20\arcsec\ of the main lens galaxy \citep{morgan}.

The situation seems different for the CB pair in PG 1115+080
and the CD pair in B0128+437.  In these cases the observed
values lie in the tail of the predicted distributions, at
around the 1\% probability level.  It may be that having two
``rare'' values among 88 image pairs is statistically
unsurprising, although it is hard to know how to quantify that
possibility because the 88 pairs are not all independent.
Alternatively, it may be that modest changes in the assumed
distributions of ellipticity, octopole moment, and shear could
raise the tail of the predicted $\Rfold$ distribution enough
to make the observed values seem less unusual.

Overall, we conclude that gross discrepancies between observed
and predicted $\Rfold$ values for the ``other'' pairs in fold
lenses indicate complex structure in the lens potential.  It is
not necessarily small-scale structure, but it is still
interesting and worth studying with detailed lens models.

\subsection{Cusp lenses}
\label{sec:obs-cusp}

\begin{figure*}[t]
\centerline{\epsfxsize=6.0in \epsfbox{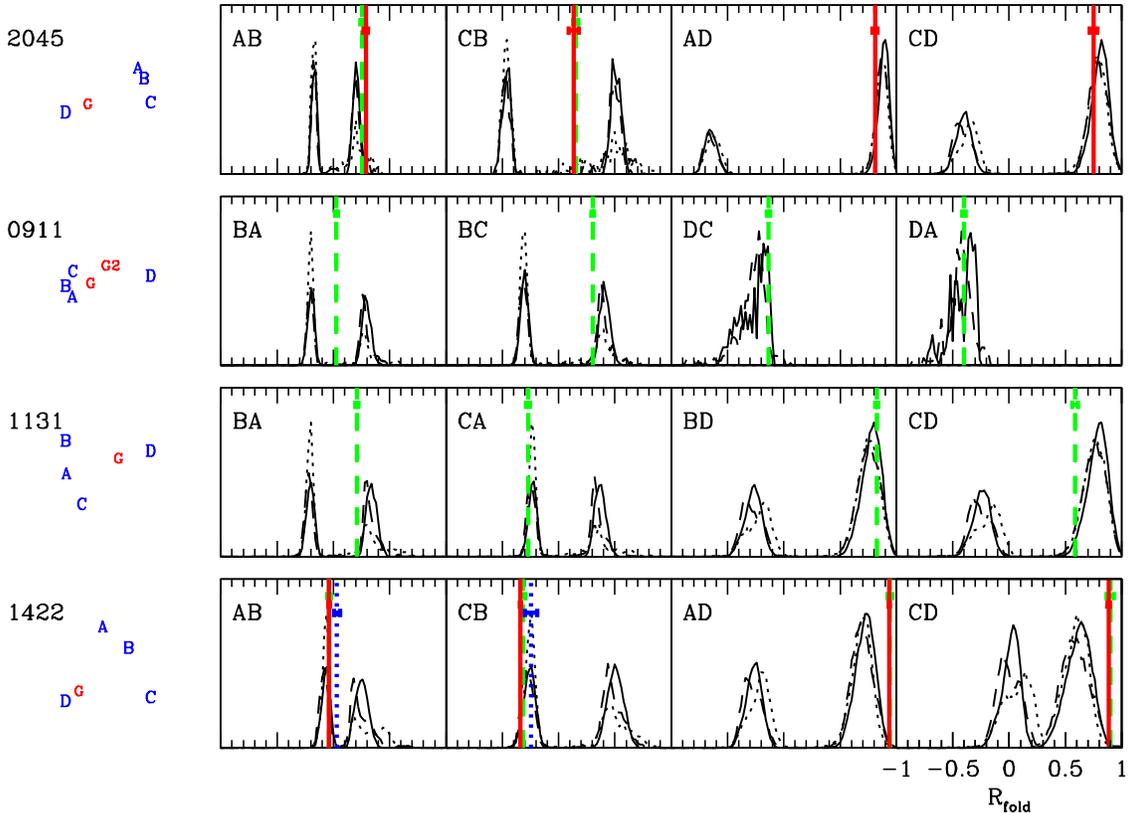}}
\caption{
Similar to Fig.~7, but for the known cusp lenses.  Again, the
vertical green lines show optical/near-infrared values of
$\Rfold$, while the vertical red lines show radio values.  For
B1422+231, the vertical blue lines show mid-infrared values of
$\Rfold$ \citep{chibaIR}.
}\label{fig:f10}
\end{figure*}

A cusp configuration has a close triplet of images that can
be thought of as two close pairs.  As discussed in
\refsecs{gen-all}{sims-res}, we exspect the predicted $\Rfold$
distribution for each pair to have two narrow and well-separated
peaks.  This constitutes a sort of ``fold relation'' for cusp
lenses, which might help us better understand flux ratio anomalies
in these systems.

Figure~10 compares the observed and predicted $\Rfold$
distributions for the four observed cusp lenses.  Image pairs
like BA in RX J0911+0551 and BA in RX J1131$-$1231, which have
observed $\Rfold$ values lying between and far from the two
predicted peaks, appear to indicate anomalies.  In one sense
these conclusions are not new, because these anomalies had
already been identified through violations of the cusp relation
(see Paper I).  However, the fold relation can help us determine
which of the three images is most anomalous.  In RX J0911+0551,
the violation of the fold relation is stronger in the BA pair
than in the BC pair, so we infer that image A is probably the
one most affected by small-scale structure.  Similar reasoning
leads to the conclusion that image B in RX J1131$-$1231 is the
most anomalous.  There could in fact be more than one perturbed
image (see \citealt{dobler} for examples among other lenses), but
the important point is that the fold relation can suggest which
image is most anomalous --- a distinction that could not be made
by the cusp relation.

The CB pair in B2045+265 illustrates a curious aspect of this
analysis.  For this lens, simulations without octopole modes
(using the J{\o}rgensen data) predict an $\Rfold$ distribution
consisting of two narrow peaks far from the observed values.
However, in simulations that include octopole modes (using the
Bender or Saglia data), there is a small but finite probability
for $\Rfold$ to lie between the two peaks.  In this case, the
possibility that octopole modes may be present limits our
ability to declare that the fold relation is violated.  This
result is surprising because B2045+265 shows a very strong
violation of the cusp relation, even when octopole modes are
considered (see Paper I).  The difference must be that the cusp
relation considered three images simultaneously, while the fold
relation considers them in two separate pairs.

Finally, we remark that the fold relation does not indicate
anomalies among three cusp images in B1422+231.  This is
consistent with our conclusion from Paper I that generic
magnification relations do not identify anomalies in this
lens, even though detailed lens modeling suggests that it is
indeed anomalous \citep{MS,bradac,MZ,dobler}.
\citet{bradac,bradac2} claim that the challenge for smooth lens
models is not just the relative brightnesses of images A, B,
and C, but also the faintness of image D.  \citet{dobler} were
able, though, to find an acceptable model under the hypothesis
that only image A is perturbed by small-scale structure.  We
conclude that the nature of the anomaly in B1422+231 is not yet
clear, and generic magnification relations are not adequate for
understanding this system.

\subsection{Cross lenses}
\label{sec:obs-cross}

\begin{figure*}[t]
\centerline{\epsfxsize=6.0in \epsfbox{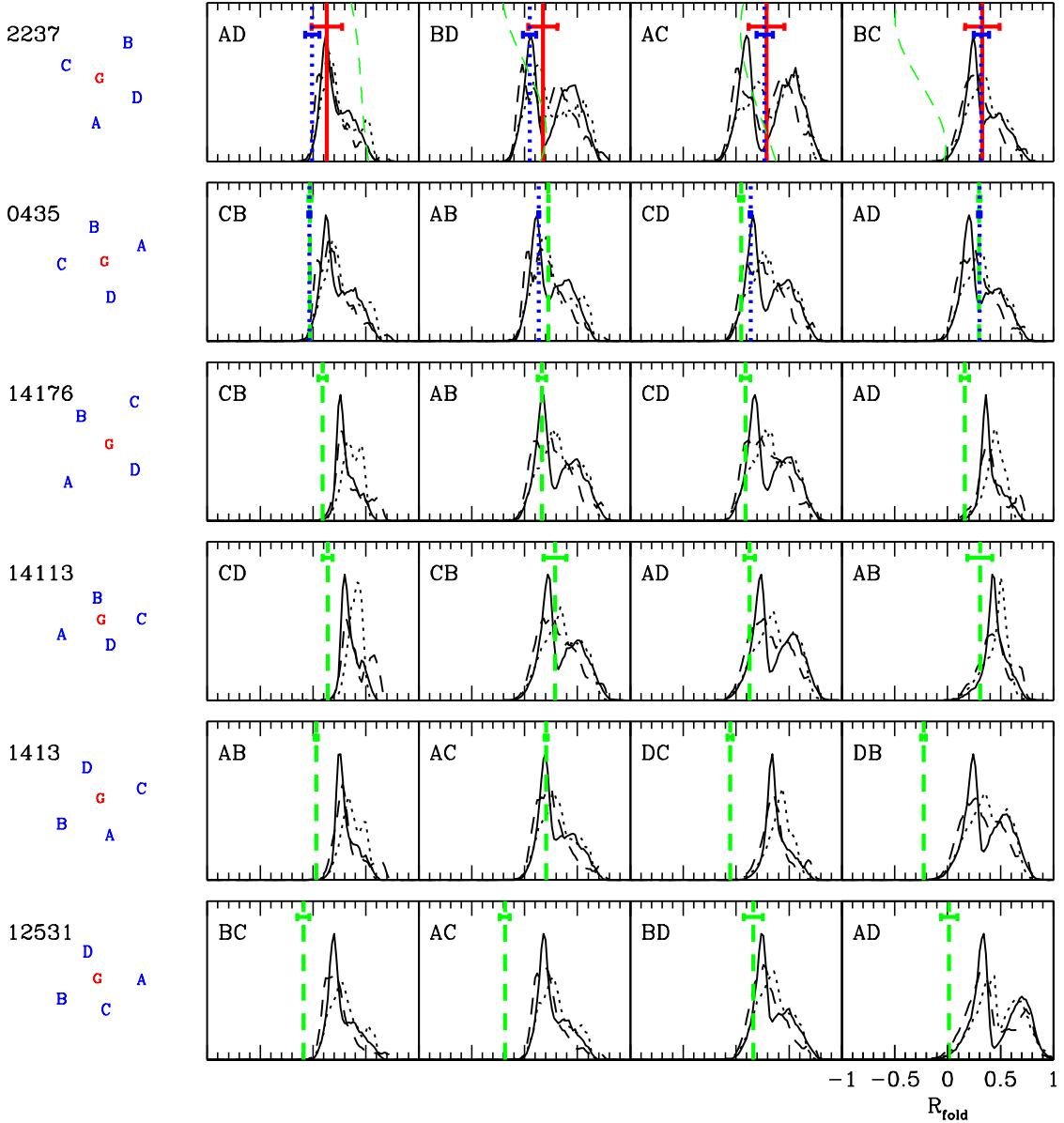}}
\caption{
Similar to Fig.~7, but for the known cross lenses.  For
Q2237$+$0305, the vertical blue lines show mid-infrared values
of $\Rfold$ \citep{agol}; also, the green lines bend to represent
the time variability in the optical flux ratios \citep{wozniak},
with time running vertically.  For HE 0435$-$1223, the vertical
blue lines indicate data from optical emission line flux ratios
\citep{wisotzki0435-2}.
}\label{fig:f11}
\end{figure*}

The image pairs in cross lenses are not close pairs, but for
completeness we still consider them in the context of the
fold relation.  Figure~11 compares the observed and predicted
$\Rfold$ values for the six known cross lenses.  The predicted
distributions are all broad and centered at some positive value
of $\Rfold$ (also see Fig.~5 and \refsec{sims-res}).  There
are several cases of disagreement, which can be understood as
follows.  First, in Q2237+0305 microlensing perturbs the flux
ratios, and in fact causes them to change with time \citep{wozniak}.

Next, in HE 0435$-$1223, HST 14113+5211, and HST 12531$-$2914,
several of the observed $\Rfold$ values lie to the left of the
predicted distributions.  However, our ellipsoid+shear lens
models fit the flux ratios fairly well, provided that the shear
is allowed to be moderately large ($\gamma = 0.13$ for
HE 0435$-$1223, and $\gamma \sim 0.3$ for the other two).  Such
large shears are rare in the distribution used for our Monte
Carlo simulations, which is why the predicted $\Rfold$
distributions in Figure~11 do not extend down to the observed
values.  They are not unreasonable, though, for lenses that
lie in complex environments like groups or clusters, which is
probably the case for all three of these lenses
\citep[see][]{wittmao,fischer,morgan0435}.  In other words, the
discrepancies in Figure~11 for these three lenses indicate that
there is ``interesting'' structure in the lens potential, but
in this case it is probably structure in the environment of the
lens as opposed to small-scale structure.

Finally, in HST 14176+5226 and HST 1413+117 there are
discrepancies between the data and predictions that are not
fully explained by a large shear.  The fold relation cannot
provide strong conclusions here, but it does suggest that these
two systems deserve further study.

A striking general feature of Figure~11 is that nearly all
the weight in the predicted probability distributions lies at
$\Rfold>0$, while some of the observed $\Rfold$ values are
negative.  We have argued that several of the negative observed
values can be explained by large shears, which are absent from
our Monte Carlo simulations.  Thus, as a rule of thumb it appears
that a negative $\Rfold$ value in a cross lens may indicate that
there is a significant environmental contribution to the lens
potential.

\section{Conclusions}
\label{sec:concl}

When the source in a four-image gravitational lens system lies
sufficiently close to a fold caustic, the two images that straddle
the fold critical curve should be mirror images of each other,
and the dimensionless flux combination
$\Rfold \equiv (F_{\rm min}-F_{\rm sad})/(F_{\rm min}+F_{\rm sad})$
should vanish.  A violation of this ``fold relation'' in an
observed lens is thought to indicate that the lens galaxy has
significant structure on scales smaller than the separation
between the two close images.  The fold relation may therefore
join the cusp relation as an important model-independent method
for identifying flux ratio anomalies that indicate small-scale
structure.

We have learned, though, that the fold relation is more subtle
and rich than was previously realized.  The ideal fold relation
$\Rfold \to 0$ holds only when the source is asymptotically
close to a fold caustic.  In more realistic situations, we find
$\Rfold \propto u^{1/2} \propto d_1$ where $u$ is the distance
of the source from the caustic, while $d_1$ is the distance
between the two close images.  In other words, $\Rfold$ goes to
zero fairly slowly as the source approaches the caustic, which
means that $\Rfold \ne 0$ might just indicate that the source
sits a finite distance from the caustic.  (For comparison, the
cusp relation has a more rapid dependence $\Rcusp \propto d^2$;
see Paper I.)  If we seek to use the fold relation to identify
flux ratio anomalies that indicate small-scale structure, then
we must understand how much $\Rfold$ can deviate from zero just
because of the finite offset of the source.

This is where we find our most startling result:  $\Rfold$ is
determined not just by the distance of the source from the
caustic, but also by the location of the source {\em along} the
caustic.  If we write $\Rfold = \Afold\,d_1 + \ldots$, then the
coefficient $\Afold$ varies enormously around the caustic.  The
problem is that the location of the source along the caustic,
and hence the value of $\Afold$, is not directly observable.
Consequently, it is no simple matter to say how large $\Rfold$
must get before we can infer the presence of small-scale structure.
Fortunately, the placement of the source along the caustic is
encoded in the image configuration: not in the separation $d_1$
between the close pair of images, but rather in the distance
$d_2$ to the next nearest image.  (For example, a source near
a fold but not near a cusp leads to $d_1 \ll d_2 \sim \Rein$,
while a source near a cusp leads to $d_1 \sim d_2 \ll \Rein$.)
We may still be able to predict the range of $\Rfold$ possible
for a smooth lens potential, but only if we consider $d_2$ as
well as $d_1$.

This general understanding allows us to develop a general
method for using the fold relation to search for flux ratio
anomalies in real lens systems.  We postulate a reasonable
and realistic population of smooth lens potentials containing
ellipticity, octopole modes, and tidal shear, and use Monte
Carlo simulations of four-image lenses produced by these lens
potentials to derive the conditional probability distribution
$p(\Rfold | d_1, d_2)$ for $\Rfold$ at fixed $d_1$ and $d_2$
(strictly speaking, fixed $d_1/\Rein$ and $d_2/\Rein$).  We
can then compare the observed value of $\Rfold$ for a real
lens to the corresponding predicted distribution to determine
whether the data are consistent with lensing by a smooth
potential.  In making our predictions, we actually consider
three different galaxy populations in order to understand how
our results depend on assumptions about what constitutes a
``reasonable and realistic'' lens population.

The fold relation applies most directly to the close pair of
images in a lens with a fold configuration.  Among the 12 known
fold lenses, we find evidence for five violations:
the optical (but not radio) fluxes in B0712+472;
the optical fluxes in SDSS 0924+0219;
the optical fluxes in PG 1115+080;
the radio fluxes in B1555+375;
and the radio fluxes in B1933+503.
The optical anomalies in B0712+472 and SDSS 0924+0219 were
already known from violations of the cusp relation (see Paper I),
but it is valuable to see them identified by the fold relation
as well.  The optical anomaly in PG 1115+080 is not quite as
secure as the others: the confidence level is 99.2\% for two
of our three sets of predictions, but only 96.1\% for one that
includes fairly strong octopole modes.  Detailed lens modeling
of PG 1115+080 suggests that multipole modes cannot provide an
acceptable explanation of the flux ratio anomaly \citep{KD,yoo},
and while that conclusion is more model-dependent than ours, it
does suggest that PG 1115+080 is indeed anomalous.  The radio
anomalies in B1555+375 and B1933+503 are newly revealed by the
fold relation.

We believe that fold flux ratio anomalies provide robust and
model-independent evidence for small-scale structure, for two
reasons.  First, the identification of the anomalies involves
a {\em local} analysis of the lens mapping, and thus relies only
on local properties of the lens potential.  This is precisely
what we want in an analysis aimed at revealing small-scale,
local structure.  Second, we have explicitly shown that (apart
from PG 1115+080) our conclusions do not change if we modify
the parameter distributions that define our realistic galaxy
population.

Based on our detailed analyses of individual fold lenses, we
can extract a few rules of thumb.  Since the sources in real
lenses always lie a finite distance from a caustic, $\Rfold$
values in the range $0 \lesssim \Rfold \lesssim 0.2$ are
predicted to be quite common and probably do {\em not} indicate
flux ratio anomalies.  When the separation between the two
close images is small ($d_1/\Rein \lesssim 0.4$), the
dependence on $d_2$ is not very strong; all that really matters
is having $d_2$ be large enough for the image configuration to
be identified as a fold.  In this case, it appears that large
values $\Rfold \gtrsim 0.2$ can reveal candidate anomalies.
However, the fact that only one of four observed lenses with
$\Rfold \approx 0.2$ is anomalous provides a strong reminder
that a full and careful analysis of the fold relation must be
done before drawing conclusions about anomalies.  Finally, as
$d_1$ increases, so too does the value of $\Rfold$ required
to indicate an anomaly, and rules of thumb about the fold
relation cease to be valid.

One final rule of thumb is that our smooth lens potentials
almost always predict $\Rfold > 0$ for fold image pairs,
indicating an asymmetry such that minimum images are generally
expected to be brighter than saddle images in fold pairs.
This point probably has implications for the prediction that
substructure affects minima and saddles differently, tending
to amplify minima but suppress saddles
\citep{MM,SW,analytics,bradac2}, and for the observation that
anomalous minima seem to be too bright while anomalous saddles
seem to be too faint \citep{KD,dobler}.  However, these issues
are not yet fully understood, and further study is needed.

Our full analysis of the fold relation also allows us to
apply it to the two close image pairs in a cusp lens.  This
application is more subtle because the analysis underlying the
fold relation breaks down near a cusp caustic.  Nevertheless,
the predicted $\Rfold$ distribution for smooth lenses is bimodal
with two narrow and well separated peaks, which constitutes a
sort of fold relation that can be used to evaluate observed
$\Rfold$ values.  We find that applying this fold relation to
cusp lenses does not reveal any new anomalies beyond those that
were identified by the cusp relation (Paper I).  However, it
may help us understand which of the three images is anomalous
(a distinction that the cusp relation cannot make).  For example,
it appears that the strongest anomaly in RX J0911+0551 is probably
in image A, while the strongest anomaly in RX J1131$-$1231 is
probably in image B.  We take these conclusions less as definite
statements and more as interesting suggestions to be examined
with detailed lens models.  One curious qualitative result is
that, in cusp lenses, the fold relation appears to be more
sensitive than the cusp relation to octopole modes.  This fact
limits our ability to find a clear violation of the fold relation
in B2045+265, even though this lens has a very strong violation
of the cusp relation.

Finally, we can also use our full understanding to apply the
fold relation to all minimum/saddle pairs in all four images
lenses, regardless of how close the pairs are.  We must be very
careful to remember that when $d_1 \gtrsim \Rein$ we are no
longer performing a {\em local} analysis of the lens mapping, so
we cannot claim to draw any model-independent conclusions about
small-scale structure.  Nevertheless, it is still interesting to
determine which lenses seem to be inconsistent with lensing by
an isothermal ellipsoid perturbed by octopole modes and moderate
shear.  Among fold lenses that do not have fold anomalies, we
find that B0128+437, HE 0230$-$2130, SDSS 1004+4112, and
WFI 2033$-$4723 all have discrepancies between the data and
predictions for other image pairs.\footnote{Actually, the
discrepancy in HE 0230$-$2130 is seen in the fold pair, but
it can be attributed to the presence of a second lens galaxy
so we prefer to discuss it here rather than among the fold
flux ratio anomalies.}  In three of these cases (SDSS 1004+4112,
HE 0230$-$2130, and WFI 2033$-$4723), the discrepancies are
(presumably) caused by complex structure in the environment of
the main lens galaxy.  Finally, each of the six known cross
lenses has at least one discrepant image pair.  In Q2237+0305
the discrepancy is caused by microlensing.  In three others
(HST 12531$-$2914, HST 14113+5211, and H1413+117) it may be
attributed to a large shear from a complex lens environment.
Again, we emphasize that discrepancies in the fold relation for
large-separation image pairs cannot be taken as strong evidence
for small-scale structure.  However, they can suggest that the
lens potential has some interesting and complex structure that
deserves further study.  

At this point it is worthwhile to review the lenses in which
violations the cusp and fold relations provide model-independent
evidence for small-scale structure in the lens potential:
\begin{itemize}
\item
Among four known cusp lenses, there are three anomalies:
RX J0911+0551 (optical), RX J1131$-$1231 (optical), and
B2045+265 (radio).
\item
Among 12 known fold lenses, there are five anomalies:
B0712+472 (optical but not radio), SDSS 0924+0219 (optical),
PG 1115+080 (optical), B1555+375 (radio), and B1933+503 (radio).
\end{itemize}
There may be other anomalies that are not identified by a
generic analysis, but that are revealed by detailed lens
modeling; B1422+231 is a prime example \citep{MS,bradac,MZ,dobler}.
Moreover, there may be systems among the ``discrepant'' lenses
mentioned above that in fact contain small-scale structure; a
good example is Q2237+0305, whose time variable discrepancies
are caused by microlensing \citep{wozniak}.  In other words,
our accounting represents a strict {\em lower} bound on the
number of lenses with flux ratio anomalies caused by small-scale
structure --- and makes it eminently clear that such anomalies
are quite common.

Interpreting these anomalies to place constraints on the nature
of the implied small-scale structure involves many considerations
that are beyond the scope of this paper.  No analysis of
single-epoch, single-band photometry can determine the scale of
the structure required to explain flux ratio anomalies, beyond
the idea that it must be smaller than the separation between the
images.  Time variability (as in Q2237+0305) or differences
between optical and radio flux ratios (as in B0712+472) may
indicate microlensing, although even then a much more detailed
analysis is required to determine the microlensing scale
\citep[e.g.,][and references therein]{csk2237}.  Absent such data,
it is impossible for any analysis of broad-band photometry in
individual lenses to robustly distinguish between microlensing,
millilensing, or intermediate-scale phenomenon.  All three
possibilities are interesting, but they have very different
implications for astrophysics.

Fortunately, there are excellent prospects for obtaining
additional data that can help distinguish between the different
hypotheses.  Even apart from time variability, comparisons between
optical continuum and broad-band flux ratios can establish the
scale of the small-scale structure \citep{moustakas,wisotzki0435-2,
metcalf2237,morgan,wayth,hst0924}.  Showing that minima and saddle
images are affected differently by small-scale structure might
also establish the scale \citep{MM,SW,analytics,KD,bradac2}.
These are several examples of the more general point that the
size of the source quasar provides a scale in the problem that
may help us determine the scale of structure in the lens
\citep{dobler}.  It is important to note that all of these
approaches require significant effort to obtain, analyze, and
interpret new data; studying {\em all} four-image lenses in this
much detail is not feasible.  It is therefore crucial to have a
reliable way to identify lenses that warrant further study.  The
fold relation joins the cusp relation in providing precisely the
realistic but robust method that we need for identifying flux
ratio anomalies.  As such, the two relations provide the foundation
for studies of small-scale structure in lens galaxies.

\acknowledgements

We thank Art Congdon and Greg Dobler for helpful discussions,
and the anonymous referee for insightful comments.
BSG was supported by a Menzel Fellowship from the Harvard College
Observatory.
AOP was supported by NSF grants DMS-0302812, AST-0434277, and
AST-0433809, and an MIT Martin Luther King, Jr.\ Visiting
Professorship in Physics.


\appendix

\section{Universal Relations for Folds}
\label{sec:AppA}

The generic properties of lensing near a fold caustic have been
studied before by \citet{BN}, \citet[Chapter 6]{SEF},
\citet[Chapter 9]{PLW}, and \citet{GPf}.  In this appendix we extend
the analysis to a higher order of approximation.

\subsection{Local orthogonal coordinates}

Consider the lens equation $\bby = \bx - \grad\psi(\bx)$.  If we
assume that the induced lensing map, $\bme(\bx) = \bx - \grad\psi(\bx)$,
from the lens plane to the source plane is locally stable, then the
caustics of $\bme$ must be either folds or cusps \citep[p.~294]{PLW}.
Let us focus on a fold caustic, and translate coordinates in the lens
and source planes so that the caustic passes through the origin
$\bby = \bmo$ of the light source plane, while the origin $\bx = \bmo$
of the lens plane maps into the origin of the light source plane.
By abuse of notation, we still use $\bx$ and $\bby$ to denote the
translated coordinates.

Consider a small neighborhood $N_L$ about the origin in the lens
plane, which maps to a local region $N_S$ about the origin in the
source plane.  We assume $N_S$ is sufficiently small that no
critical points outside $N_L$ are mapped into $N_S$, and there are
no cusp caustic points inside $N_S$.  In other words, the only
caustic in $N_S$ is a fold arc passing through the origin.

By Taylor expanding, we see that the Jacobian matrix of the lensing
map $\bme$ is given at the origin $\bx = \bmo$ by
\begin{equation}
  [\jac \bme](\bmo) =
    \left[ \begin{array}{cc}
      1 - 2\ha & -\hb \\
      -\hb & 1 - 2\hc
    \end{array} \right],
\end{equation}
where
\begin{equation}
  \ha = \frac{1}{2} \psi_{11} (\bmo), \qquad
  \hb =  \psi_{12} (\bmo), \qquad
  \hc =  \frac{1}{2} \psi_{22} (\bmo).
\end{equation}
The subscripts indicate partial derivatives of $\psi$ relative to
$\bx = (x_1, x_2)$.  Note that $\psi$ has no linear part (since $\bme$
maps the origin to itself).  For $\bby = \bmo$ to be fold caustic
point, the rank of $[\jac \bme](\bmo)$ must be unity, which means
that we must have $(1-2\ha)(1-2\hc) - \hb^2 = 0$ while at least one
of $(1-2\ha)$, $(1-2\hc)$, and $\hb^2$ does not vanish
\citep[p.~349]{PLW}.  Consequently, $(1-2\ha)$ and $(1-2\hc)$ cannot
both vanish.  We lose no generality by assuming $1-2\ha \neq 0$.

Now introduce the orthogonal matrix \citep[see][p.~344]{PLW}
\begin{equation}
  {\bf M} = \frac{1}{\sqrt{(1 - 2\ha)^2 + \hb^2}}
    \left[ \begin{array}{cc}
      1 - 2\ha & -\hb \\
      \hb & 1 - 2\ha
    \end{array} \right] ,
\end{equation}
and define new orthogonal coordinates in the neighborhoods $N_L$ and
$N_S$ by 
\begin{equation}
  \bt = (\theta_1, \theta_2) \equiv  {\bf M}\, \bx ,
  \qquad
  \bu = (u_1, u_2) \equiv {\bf M}\, \bby .
\end{equation}
(Note that the coordinate changes are the {\em same} in the lens
and source planes, which is different from the approach of
\citealt{SEF}, p.~185.)  Using these coordinates, \citet[p.~346]{PLW}
showed rigorously that $\bx = \bmo$ is a fold critical point if and
only if the following conditions hold:
\begin{equation} \label{eq:eq-pot-fconstraints}
  (1-2\ha)(1-2\hc) = \hb^2 , \quad
  1-2\ha \neq 0 , \quad
  \hd \equiv - \psi_{222} (\bmo) \neq 0 .
\end{equation}

{\em Remark:} The matrix ${\bf M}$ orthogonally diagonalizes
$[\jac \bme](\bmo)$.

Let us now Taylor expand the lens potential near the origin.  We
argue below that carrying the expansion to fourth order in $\bt$
is both necessary and sufficient for the precision we desire.  The
most general fourth order expansion can be written as
\citep[see][pp.~346--347]{PLW}
\begin{equation}
  \psi(\bt) =
      \frac{1}{2}(1-K)\,\theta_1^2
    + \frac{1}{2}\,\theta_2^2
    + e\,\theta_1^3
    + f\,\theta_1^2 \theta_2
    + g\,\theta_1   \theta_2^2
    + h\,           \theta_2^3
    + k\,\theta_1^4
    + m\,\theta_1^3 \theta_2
    + n\,\theta_1^2 \theta_2^2
    + p\,\theta_1   \theta_2^3
    + r\,           \theta_2^4 . \label{eq:psiexp}
\end{equation}
The zeroth order term in the potential is irrelevant, so we neglect
it.  The first order terms must vanish in order to ensure that the
origin of the lens plane maps to the origin of the source plane.
In the second order terms, the coefficients of the $\theta_1 \theta_2$
and $\theta_2^2$ terms are fixed (to 0 and $1/2$, respectively) by
the conditions that the origin is a fold critical point such that
$[\jac \bme](\bmo)$ is in diagonal form.  Note that the coefficient
$e$ of the $\theta_1^3$ term here is different from the ellipticity
parameter used in the main text.  We retain $e$ here to match the
notation used by \citet{PLW}.  The $e$ coefficient does not appear
in the main text, and the ellipticity parameter does not appear
explicitly in this appendix, so there should be little confusion.

Conventional analyses of lensing near a fold caustic have only
considered the $K$, $e$, $f$, $g$, and $h$ terms in the expansion
\citep{BN,SEF,PLW,GPf}.  However, we shall see below that some
of the other terms are significant for our analysis.

\subsection{Image positions and magnifications}

We seek to use perturbation theory \citep[e.g.,][]{bellman} to find
expansions for the image positions and magnifications that are
accurate to first order in $\bu$.  We shall work from our fourth
order expansion of the lens potential, and then verify that it is
adequate for our purposes.  For bookkeeping purposes, let us
introduce scalar parameter $\xi$ by taking $\bu \to \xi \bu$, so we
can identify terms of a given order by examining the power of $\xi$.

For the potential \eq{psiexp}, the lens equation is
\begin{eqnarray}
   \xi u_1 &=& K\,\theta_1
    - \left( 3 e\,\theta_1^2 + 2 f\,\theta_1 \theta_2 + g\,\theta_2^2 \right)
    - \left( 4 k\,\theta_1^3 + 3 m\,\theta_1^2 \theta_2
           + 2 n\,\theta_1 \theta_2^2 + p\,\theta_2^3 \right) ,
    \label{eq:lens1-u1} \\
   \xi u_2 &=& - \left( f\,\theta_1^2 + 2 g\,\theta_1 \theta_2
    + 3 h\,\theta_2^2 \right) - \left( m\,\theta_1^3
    + 2 n\,\theta_1^2 \theta_2 + 3 p\,\theta_1 \theta_2^2
    + 4 r\,\theta_2^3 \right) .
    \label{eq:lens1-u2}
\end{eqnarray}
Since the lowest order terms are linear or quadratic in $\bt$, it
is natural to postulate that the image positions can be written as
a series expansion in $\xi$ with the following form:
\begin{eqnarray}
  \theta_1 &=& \alpha_1\,\xi^{1/2} + \beta_1\,\xi + \order{\xi}{3/2} , \\
  \theta_2 &=& \alpha_2\,\xi^{1/2} + \beta_2\,\xi + \order{\xi}{3/2} ,
\end{eqnarray}
Substituting into the lens equation, we obtain:
\begin{eqnarray}
  0 &=& (\alpha_1 K)\xi^{1/2} - (3 \alpha_1^2 e + 2 \alpha_1 \alpha_2 f
    + \alpha_2^2 g - \beta_1 K + u_1)\xi + \order{\xi}{3/2} ,
    \label{eq:lens2-u1} \\
  0 &=& -(\alpha_1^2 f + 2 \alpha_1 \alpha_2 g + 3 \alpha_2^2 h + u_2) \xi
    - \left[ 2 \alpha_1 \beta_1 f
      + 2 (\alpha_1 \beta_2 + \alpha_2 \beta_1) g + 6 \alpha_2 \beta_2 h
      + \alpha_1^3 m + 2 \alpha_1^2 \alpha_2 n + 3 \alpha_1 \alpha_2^2 p
      + 4 \alpha_2^3 r \right] \xi^{3/2}
    + \order{\xi}{2} .
    \label{eq:lens2-u2}
\end{eqnarray}
It is easily understood why these equations are carried to different
orders.  Eq.~(\ref{eq:lens1-u1}) has a term that is linear in $\bt$,
which means that correction terms appear at $\order{\xi}{3/2}$.  By
contrast, in eq.~(\ref{eq:lens1-u2}) the lowest order term is
quadratic in $\bt$, and since
\begin{equation}
  \theta_i \theta_j = \alpha_i \alpha_j \xi
    + (\alpha_i \beta_j + \alpha_j \beta_i) \xi^{3/2}
    + \order{\xi}{2}
\end{equation}
we see that the correction terms only appear at $\order{\xi}{2}$.

Following perturbation theory, we can now solve for the unknowns
$\alpha_i$ and $\beta_i$ by demanding that eqs.~(\ref{eq:lens2-u1})
and (\ref{eq:lens2-u2}) be satisfied at each order in $\xi$.  We
then find that the positions of the two images can be written as
\begin{eqnarray}
  \theta_1^{\pm} &=& \frac{3 h u_1 - g u_2}{3 h K}\ \xi + \order{\xi}{3/2} , \\
  \theta_2^{\pm} &=& \mp \sqrt{\frac{-u_2}{3h}}\ \xi^{1/2}
    - \frac{3 g h u_1 - (g^2 + 2 K r) u_2}{9 h^2 K}\ \xi + \order{\xi}{3/2} .
\end{eqnarray}
Note that the distance between the two images is
\begin{equation} \label{eq:d1ser}
  d_1 = 2 \sqrt{\frac{-u_2}{3h}}\ \xi^{1/2} + \order{\xi}{3/2} .
\end{equation}

To find the magnifications of the images, we compute the Jacobian
determinant of the lens equation, and evaluate that at $\bt^{\pm}$
to obtain
\begin{equation} \label{eq:muser}
  \left(\mu^{\pm}\right)^{-1} = \pm 2 K \sqrt{-3 h u_2}\ \xi^{1/2}
    + \frac{4}{3h}\left(g^2 - 3 f h + 2 K r\right) u_2\,\xi + \order{\xi}{3/2} .
\end{equation}
This result shows that the $\pm$ labels for the two images have been
assigned such that $\mu^{+} > 0$ while $\mu^{-} < 0$.

{\em Remarks:}
(1) To check our results, we note that at lowest order we recover the
same scalings $d_1 \propto \xi^{1/2}$ and $\mu^{\pm} \propto \xi^{-1/2}$
found by previous analyses \citep{SEF,PLW,GPf}.
(2) To first order in $\xi$ the image separation and the two
magnifications depend only on the $u_2$ component of the source
position.
(3) In several places we have $\sqrt{-h u_2}$ or $\sqrt{-u_2/h}$.
In general, at least for simple lens potentials like an isothermal
ellipsoid or isothermal sphere with shear, we have $h \le 0$ all
along the caustic.  This means that only source positions with
$u_2 > 0$ lead to the production of two fold images.

At first order in the image positions and magnifications, the
presence of $r$ demonstrates that the fourth order terms in
\eq{psiexp} cannot be ignored.  At the same time, we can now
verify that going to fourth order is sufficient.  Any term of
$\mathcal{O}(\bt)^5$ in the potential would enter the lens equation
at $\mathcal{O}(\bt)^4$; that would in turn be of order $\xi^2$
or higher, which is beyond the order to which we are working.
Similarly, terms of $\mathcal{O}(\bt)^5$ in the potential would enter
$\mu^{-1}$ at $\mathcal{O}(\bt)^3$ or at least $\xi^{3/2}$.  In other
words, going to fourth order in \eq{psiexp} is both necessary and
sufficient when we seek the image positions and magnifications
correct to first order in $\xi$.

\subsection{Generic behavior of the fold relation}

From \eq{muser} we see that the two fold images have magnifications
that are equal and opposite to lowest order in $\xi$, which means
that the combination $|\mu^{+}| - |\mu^{-}|$ should approximately
vanish.  In observed lenses, the magnifications are not directly
observable but the fluxes are, so to construct a dimensionless
combination of the fluxes we define
\begin{equation}
  \Rfold \equiv
  \frac{|\mu^{+}| - |\mu^{-}|}{|\mu^{+}| + |\mu^{-}|}
  = \frac{F^{+} - F^{-}}{F^{+} + F^{-}}\ .
\end{equation}
Plugging in the series expansions for $\mu^{\pm}$, we find
\begin{equation} \label{eq:Rfold-src}
  \Rfold = \frac{2(g^2 - 3 f h + 2 K r)}{3 h K}
    \sqrt{\frac{-u_2}{3h}}\ \xi^{1/2} + \order{\xi}{} .
\end{equation}
By comparing the expansion for $d_1$ in \eq{d1ser}, we see that
we can write
\begin{equation} \label{eq:Rfold-img}
  \Rfold = \Afold\,d_1 + \order{\xi}{} ,
\end{equation}
where
\begin{equation} \label{eq:Afold}
  \Afold \equiv \frac{g^2 - 3 f h + 2 K r}{3 h K}
  = \frac{3 \psi_{122}^2 - 3 \psi_{112} \psi_{222} + \psi_{2222} (1-\psi_{11})}
    {6 \psi_{222} (1-\psi_{11})}\ .
\end{equation}
In the last equality, we have replaced the coefficients $(f,g,h,K,r)$
with their definitions in terms of derivatives of the potential;
for example,
\begin{eqnarray}
  f &=& \frac{1}{2}\,\psi_{112}
    = \frac{1}{2}\ \frac{\partial^3\psi}{\partial\theta_1^2 \partial\theta_2}\ ,
    \label{eq:fdef} \\
  g &=& \frac{1}{2}\,\psi_{122}
    = \frac{1}{2}\ \frac{\partial^3\psi}{\partial\theta_1 \partial\theta_2^2}\ ,
    \label{eq:gdef}
\end{eqnarray}
and so forth, where the derivatives are evaluated at the origin
$\bt = \bmo$.

To summarize, $\Rfold$ vanishes for a source asymptotically close
to a fold caustic.  For a source a small but finite distance away,
there is a correction term that scales as the square root of the
distance of the source from the caustic, or (more usefully) as the
separation between the two fold images.  The coefficient $\Afold$
of this linear scaling depends on properties of the lens potential
at the fold critical point.  In particular, the presence of $r$
reiterates the fact that the fourth order expansion in the potential
(eq.~\ref{eq:psiexp}) is necessary to obtain an expansion for
$\Rfold$ that is accurate at order $\xi^{1/2}$ or $d_1$.

We have verified all of the approximations in this appendix by
comparing them to exact numerical solutions of the lens equation
obtained with the software by \citet{lenscode}.

\section{Data for the Observed Four-Image Lenses}
\label{sec:AppB}

In this appendix we summarize the observational data that we use
for all of the observed four-image lenses; this text complements
the data values given in Table~1.  For each lens system, we need
the relative positions of the lensed images in order to measure
the separations $d_1$ and $d_2$, and also to use as constraints
on lens models for determining the Einstein radius $\Rein$.  The
position of the lens galaxy (or galaxies), if available, is also
valuable for the lens modeling.  We also need the flux ratios
between the images in order to determine $\Rfold$.  We consider
radio and optical/near-infrared flux ratios separately, because
they are believed to correspond to very different source sizes and
provide very different information about small-scale structure in
the lens potential \citep[see, e.g.,][]{DK,dobler}.  At
optical/near-infrared wavelengths, we examine the colors of the
images to detect (and correct for) any differential extinction that
may be present.  \citet{falco} carried out a detailed version of
this analysis for a sample of lenses that includes seven that we
consider, and we use their results where available.  If there is
no evidence for differential extinction, we combine data from
different passbands using a weighted average.  At radio wavelengths,
we again examine the wavelength dependence of the flux ratios to
determine that there are no significant electromagnetic effects.
If there are other flux measurements, such as in optical broad
emission lines or at mid-infrared wavelengths, we use those as
well (as discussed below).

{\bf 0047$-$2808:}  This is a quadruply imaged system
\citep{warren96,warren99}, but its lack of pointlike images
means that it requires special modeling techniques
\citep[e.g.,][]{wayth0047}, and that it is probably not very
interesting for the analysis of small-scale structure anyway.  We
do not include it in our sample.

{\bf B0128+437:}  For the image positions, we use the radio
astrometry from \citet{phillips}.  For the radio fluxes, we use
the mean and scatter in $\Rfold$ from 41 epochs of MERLIN
monitoring by \citet{koopmans}.  The monitoring shows no evidence
for time dependence.

{\bf HE 0230$-$2130:}  We use the optical HST astrometry from
CASTLES.  There are two lens galaxies; we include both of them in
lens models, taking their observed positions as constraints but
optimizing their masses.  For the optical fluxes of the lensed
images, we use the BRI data for images A, B, and C from
\citet{wisotzki0230}.  The colors are consistent with no
differential extinction.  (Image D is not well separated from
galaxy G2 in the \citeauthor{wisotzki0230} data, so we do not
consider it.)

{\bf MG 0414+0534:}  We use the optical HST astrometry from
CASTLES.  We include the satellite galaxy near the lens galaxy
\citep[``object X''][]{SchechterMoore} in lens models.  For the
optical image fluxes, we use the extinction-corrected flux ratios
from \citet{falco}.  For the radio fluxes, we use the high-resolution
VLBI data from \citet{trotter}.  Those observations resolve each
image into four subcomponents; the $\Rfold$ values are similar for
the different subcomponents, so we take the weighted average.  The
radio flux ratios are constant in time to 1--3\% \citep{moore}.

{\bf HE 0435$-$1223:}  We use the optical HST astrometry from
CASTLES.  For the optical broad-band fluxes, we use the $gri$
data from \citet{wisotzki0435-1}.  \citet{wisotzki0435-2} also
report emission line fluxes; we take the mean and scatter in
$\Rfold$ from the \ion{C}{4} and \ion{C}{3}] lines.  There is no
evidence for wavelength dependence in the broad-band flux ratios,
and the spectra of the different images have identical spectral
slopes, so there does not appear to be any differential extinction.
Image D appeared to vary by 0.07 mag between the two sets of
observations, which may imply evidence for microlensing.

{\bf B0712+472:}  We use the optical HST astrometry and photometry
from CASTLES.  The values of $\Rfold$ differ slightly in the V,
I, and H bands, but within the measurement uncertainties; hence
there is no evidence for differential extinction.  For the radio
fluxes, we use the mean and scatter in $\Rfold$ from MERLIN
monitoring by \citet{koopmans}.  There is evidence for time
dependence in the radio fluxes.

{\bf RX J0911+0551:}  We use the optical HST astrometry and
photometry from CASTLES.  The lens galaxy has a satellite galaxy,
which we include in lens models.  The image flux ratios vary with
wavelength in a manner that is consistent with differential
extinction, so we correct for extinction using a redshifted
$R_V=3.1$ extinction curve from \citet[also see Paper I]{cardelli}.

{\bf SDSS 0924+0219:}  \citet{hst0924} report image positions,
broad-band flux ratios, and broad emission line flux ratios from
HST observations.  We use the weighted average of the V and I
broad-band flux ratios.  We use the emission line fluxes with 5\%
uncertainties, which is probably conservative.  The best color
information comes from $gri$ data by \citet{inada0924}, which are
consistent with no differential extinction.

{\bf SDSS 1004+4112:}  This lens is produced by a cluster rather
than a single galaxy \citep{oguri1004}, but we can still treat it
with our formalism.  We use ground-based $griz$ data from
\citet{oguri1004}, and HST/I data from \citet{inada1004}.  There
is no evidence for differential extinction.  \citet{richards}
claimed to observe microlensing of the broad emission lines in
image A, but the level of variability in the continuum is not yet
known.

{\bf PG 1115+080:}  We use the HST astrometry and photometry from
CASTLES, and the mid-infrared flux ratios from \citet{chibaIR}.
\citet{falco} find that the VIH data are consistent with no
differential extinction.  In the lens models, we explicitly
include the group of galaxies surrounding the lens
\cite[see][]{kk1115,impey1115}.

{\bf RX J1131$-$1231:}  We use the ground-based astrometry and
photometry from \citet{sluse}.  They report two epochs of V data
and one epoch of R.  The colors are consistent with no differential
extinction.  The total flux varied between the two epochs, but the
flux ratios remained constant.

{\bf HST 12531$-$2914:}  We use the HST astrometry and photometry
from \citet{ratnatunga} and CASTLES.  \citet{falco} find that the
V$-$I colors are consistent with no extinction (within the noise).

{\bf HST 14113+5211:}  We use the HST astrometry and photometry
from \citet{fischer} and CASTLES.  There is some scatter among
the values of $\Rfold$ obtained from V-, R-, and I-band data, but
the scatter is within the (fairly large) measurement uncertainties.

{\bf H1413+117:}  We use the HST astrometry from CASTLES.  For the
optical fluxes, we use the extinction-corrected flux ratios from
\citet{falco}.

{\bf HST 14176+5226:}  We use the HST astrometry and photometry
from \citet{ratnatunga}.  \citet{falco} find that the colors are
consistent with no differential extinction.

{\bf B1422+231:}  We use the radio data from \citet{patnaik}.  The
radio fluxes are basically constant in time \citep{patnaik2}.  For
the optical fluxes, we use the extinction-corrected flux ratios
from \citet{falco}.  We also use the mid-infrared flux ratios
between images A, B, and C (image D was not detected) from
\citet{chibaIR}.

{\bf B1555+375:}  We use the radio data from \citet{marlow}.  The
data from radio monitoring by \citet{koopmans} yield similar results,
but have larger formal errors.

{\bf B1608+656:}  \citet{fassnacht1608} monitored the radio fluxes,
measured the time delays, and determined the delay-corrected
magnification ratios; we take the mean and scatter in $\Rfold$ from
their three seasons of data.  There are two lens galaxies; we model
the system using data from \citet{koopmans1608}.

{\bf B1933+503:}  There are ten lensed images associated with three
different sources.  We use all of the images in lens modeling,
following \citet{cohn}.  However, for the fold analysis we use
only the fold quad consisting of images 1/3/4/6.  For the radio
fluxes, we first take the mean and scatter from 8.4 GHz monitoring
by \citet{biggs}, and then combine that in weighted average with
measurements at other wavelengths by \citet{sykes}.

{\bf WFI 2026$-$4536:}  We use the optical data from \citet{morgan}.
We use all available data in which the images are resolved:
$ugriHK_s$ plus HST/F160W for images B and C; and $iHK_s$ plus
HST/F160W for images A$_1$ and A$_2$.  There is some wavelength
dependence that may suggest differential extinction or microlensing,
the current data are inconclusive.  We simply take the mean and
scatter in $\Rfold$ from all of the data.

{\bf WFI 2033$-$4723:}  We use the optical data from \citet{morgan}.
For the optical broad-band flux ratios, we use all available data
in which the images are resolved: $ugri$ for images B and C; and
$ri$ for images A$_1$ and A$_2$.  \citeauthor{morgan} also report
emission line flux ratios; we take the weighted average of $\Rfold$
from the \ion{C}{4}, \ion{C}{3}], and \ion{Mg}{2} lines.

{\bf B2045+265:}  We use the radio positions from
\citet{fassnacht2045}.  For the radio fluxes, we combine various
measurements by \citet{fassnacht2045} and monitoring by
\citet{koopmans}, and take the mean and scatter in $\Rfold$.  For
the optical fluxes, we use HST data from CASTLES for images A, B,
and C (image D was not detected).  The VIH colors are consistent
with no differential extinction.

{\bf Q2237+0305:}  We use HST astrometry from CASTLES.  For the
broad-band optical fluxes, we use the microlensing light curves
from \citet{wozniak}.  We correct for differential extinction using
the reddening deduced by \citet{falco}.  For the radio fluxes, we
use the data from \citet{falco2237}.  In addition, \citet{agol}
report mid-infrared flux ratios measured at 8.9 $\mu$m and
11.7 $\mu$m.



\clearpage
\LongTables

\begin{deluxetable}{ccrrrrl}
\tablewidth{0pt}
\tablecaption{Lens Data}
\tablehead{
 &
 \colhead{Image Pair} &
 \colhead{$d_1$} &
 \colhead{$\Rfold$ (optical)} &
 \colhead{$\Rfold$ (radio)} &
 \colhead{$\Rfold$ (other)} &
 \colhead{References}
}
\startdata
B0128$+$437     &          AB &  0.14 &                  & $ 0.263\pm0.023$ &                  & \citet{koopmans} \\
$\Rein =  0.20$ &          AD &  0.27 &                  & $ 0.328\pm0.028$ &                  & \\
fold            &          CD &  0.42 &                  & $ 0.014\pm0.042$ &                  & \\
                &          CB &  0.50 &                  & $-0.058\pm0.037$ &                  & \\
\tableline
HE0230$-$2130   &          AB &  0.74 & $ 0.000\pm0.008$ &                  &                  & \citet{wisotzki0230} \\
$\Rein =  0.82$ &          CD &  1.46 &                  &                  &                  & \\
fold            &          AD &  1.64 &                  &                  &                  & \\
                &          CB &  1.65 & $-0.289\pm0.007$ &                  &                  & \\
\tableline
MG0414$+$0534   &  A$_1$A$_2$ &  0.41 & $-0.024\pm0.038$ & $ 0.085\pm0.002$ &                  & \citet{falco}, \\
$\Rein =  1.08$ &      BA$_2$ &  1.71 & $-0.500\pm0.043$ & $-0.477\pm0.004$ &                  & \citet{trotter} \\
fold            &      A$_1$C &  1.96 & $ 0.739\pm0.015$ & $ 0.736\pm0.003$ &                  & \\
                &          BC &  2.13 & $ 0.400\pm0.046$ & $ 0.323\pm0.007$ &                  & \\
\tableline
HE0435$-$1223   &          CB &  1.53 & $-0.029\pm0.014$ &                  & $-0.035\pm0.010$ & \citet{wisotzki0435-1}, \\
$\Rein =  1.18$ &          AB &  1.59 & $ 0.226\pm0.004$ &                  & $ 0.136\pm0.005$ & \citet{wisotzki0435-2} \\
cross           &          CD &  1.85 & $ 0.049\pm0.019$ &                  & $ 0.137\pm0.007$ & \\
                &          AD &  1.88 & $ 0.299\pm0.012$ &                  & $ 0.300\pm0.011$ & \\
\tableline
B0712$+$472     &          AB &  0.17 & $ 0.519\pm0.052$ & $ 0.085\pm0.036$ &                  & \citet{koopmans}, \\
$\Rein =  0.68$ &          CB &  0.91 & $-0.123\pm0.075$ & $-0.337\pm0.051$ &                  & CASTLES \\
fold/cusp       &          CD &  1.18 & $ 0.361\pm0.062$ & $ 0.672\pm0.120$ &                  & \\
                &          AD &  1.25 & $ 0.636\pm0.062$ & $ 0.848\pm0.060$ &                  & \\
\tableline
RXJ0911$+$0551  &          BA &  0.48 & $ 0.027\pm0.013$ &                  &                  & CASTLES \\
$\Rein =  0.95$ &          BC &  0.62 & $ 0.303\pm0.012$ &                  &                  & \\
cusp            &          DC &  2.96 & $-0.137\pm0.016$ &                  &                  & \\
                &          DA &  3.08 & $-0.400\pm0.014$ &                  &                  & \\
\tableline
SDSS0924$+$0219 &          AD &  0.69 & $ 0.873\pm0.002$ &                  & $ 0.821\pm0.012$ & \citet{hst0924} \\
$\Rein =  0.87$ &          AC &  1.18 & $ 0.483\pm0.003$ &                  & $ 0.696\pm0.019$ & \\
fold            &          BD &  1.46 & $ 0.751\pm0.002$ &                  & $ 0.593\pm0.023$ & \\
                &          BC &  1.53 & $ 0.149\pm0.002$ &                  & $ 0.363\pm0.031$ & \\
\tableline
SDSS1004$+$4112 &          AB &  3.73 & $ 0.194\pm0.015$ &                  &                  & \citet{oguri1004}, \\
$\Rein =  6.91$ &          DB & 11.44 & $-0.512\pm0.017$ &                  &                  & \citet{inada1004} \\
fold            &          AC & 11.84 & $ 0.401\pm0.011$ &                  &                  & \\
                &          DC & 14.38 & $-0.312\pm0.016$ &                  &                  & \\
\tableline
PG1115$+$080    &  A$_1$A$_2$ &  0.48 & $ 0.215\pm0.011$ &                  & $ 0.036\pm0.032$ & CASTLES, \\
$\Rein =  1.03$ &      A$_1$B &  1.67 & $ 0.722\pm0.009$ &                  & $ 0.724\pm0.104$ & \citet{chibaIR} \\
fold            &          CB &  1.99 & $ 0.214\pm0.019$ &                  & $ 0.135\pm0.234$ & \\
                &      CA$_2$ &  2.16 & $-0.445\pm0.011$ &                  & $-0.632\pm0.060$ & \\
\tableline
RXJ1131$-$1231  &          BA &  1.19 & $ 0.209\pm0.013$ &                  &                  & \citet{sluse} \\
$\Rein =  1.81$ &          CA &  1.26 & $-0.272\pm0.019$ &                  &                  & \\
cusp            &          BD &  3.14 & $ 0.824\pm0.012$ &                  &                  & \\
                &          CD &  3.18 & $ 0.587\pm0.026$ &                  &                  & \\
\tableline
HST12531$-$2914 &          BC &  0.77 & $-0.092\pm0.057$ &                  &                  & \citet{ratnatunga}, \\
$\Rein =  0.55$ &          AC &  0.78 & $-0.187\pm0.046$ &                  &                  & CASTLES \\
cross           &          BD &  0.91 & $ 0.164\pm0.089$ &                  &                  & \\
                &          AD &  1.02 & $ 0.015\pm0.078$ &                  &                  & \\
\tableline
HST14113$+$5211 &          CD &  1.13 & $ 0.138\pm0.049$ &                  &                  & \citet{fischer}, \\
$\Rein =  0.83$ &          CB &  1.38 & $ 0.287\pm0.109$ &                  &                  & CASTLES \\
cross           &          AD &  1.41 & $ 0.128\pm0.049$ &                  &                  & \\
                &          AB &  1.42 & $ 0.305\pm0.116$ &                  &                  & \\
\tableline
H1413$+$117     &          AB &  0.76 & $ 0.031\pm0.016$ &                  &                  & \citet{falco} \\
$\Rein =  0.56$ &          AC &  0.87 & $ 0.205\pm0.015$ &                  &                  & \\
cross           &          DC &  0.91 & $-0.056\pm0.023$ &                  &                  & \\
                &          DB &  0.96 & $-0.229\pm0.022$ &                  &                  & \\
\tableline
HST14176$+$5226 &          CB &  1.73 & $ 0.088\pm0.040$ &                  &                  & \citet{ratnatunga} \\
$\Rein =  1.33$ &          AB &  2.09 & $ 0.163\pm0.040$ &                  &                  & \\
cross           &          CD &  2.13 & $ 0.089\pm0.043$ &                  &                  & \\
                &          AD &  2.13 & $ 0.164\pm0.040$ &                  &                  & \\
\tableline
B1422$+$231     &          AB &  0.50 & $-0.038\pm0.018$ & $-0.038\pm0.007$ & $ 0.031\pm0.027$ & \citet{falco}, \\
$\Rein =  0.76$ &          CB &  0.82 & $-0.317\pm0.020$ & $-0.339\pm0.006$ & $-0.245\pm0.055$ & \citet{patnaik}, \\
cusp            &          AD &  1.25 & $ 0.942\pm0.019$ & $ 0.936\pm0.006$ &                  & \citet{chibaIR} \\
                &          CD &  1.29 & $ 0.898\pm0.032$ & $ 0.884\pm0.011$ &                  & \\
\tableline
B1555$+$375     &          AB &  0.09 &                  & $ 0.274\pm0.003$ &                  & \citet{marlow}, \\
$\Rein =  0.24$ &          CB &  0.35 &                  & $-0.084\pm0.004$ &                  & \citet{koopmans} \\
fold            &          AD &  0.40 &                  & $ 0.858\pm0.006$ &                  & \\
                &          CD &  0.42 &                  & $ 0.725\pm0.010$ &                  & \\
\tableline
B1608$+$656     &          AC &  0.87 &                  & $ 0.321\pm0.006$ &                  & \citet{fassnacht1608} \\
$\Rein =  0.77$ &          BC &  1.51 &                  & $-0.016\pm0.002$ &                  & \\
fold            &          AD &  1.69 &                  & $ 0.706\pm0.004$ &                  & \\
                &          BD &  2.00 &                  & $ 0.486\pm0.004$ &                  & \\
\tableline
B1933$+$503     &         4-3 &  0.46 &                  & $ 0.656\pm0.007$ &                  & \citet{sykes}, \\
$\Rein =  0.49$ &         4-6 &  0.63 &                  & $ 0.637\pm0.007$ &                  & \citet{biggs} \\
fold            &         1-3 &  0.90 &                  & $ 0.143\pm0.014$ &                  & \\
                &         1-6 &  0.91 &                  & $ 0.111\pm0.013$ &                  & \\
\tableline
WFI2026$-$4536  &  A$_1$A$_2$ &  0.33 & $ 0.181\pm0.043$ &                  &                  & \citet{morgan} \\
$\Rein =  0.65$ &      A$_1$C &  0.83 & $ 0.626\pm0.015$ &                  &                  & \\
fold            &          BC &  1.19 & $-0.431\pm0.014$ &                  &                  & \\
                &      BA$_2$ &  1.28 & $ 0.096\pm0.011$ &                  &                  & \\
\tableline
WFI2033$-$4723  &  A$_1$A$_2$ &  0.72 & $ 0.219\pm0.010$ &                  & $ 0.174\pm0.099$ & \citet{morgan} \\
$\Rein =  1.06$ &      A$_1$C &  1.54 & $ 0.330\pm0.007$ &                  & $ 0.056\pm0.074$ & \\
fold            &      BA$_2$ &  2.01 & $-0.072\pm0.012$ &                  & $-0.042\pm0.091$ & \\
                &          BC &  2.13 & $ 0.077\pm0.044$ &                  & $-0.161\pm0.057$ & \\
\tableline
B2045$+$265     &          AB &  0.28 & $ 0.255\pm0.017$ & $ 0.287\pm0.020$ &                  & \citet{fassnacht2045}, \\
$\Rein =  1.13$ &          CB &  0.56 & $ 0.153\pm0.023$ & $ 0.133\pm0.045$ &                  & CASTLES \\
cusp            &          AD &  1.91 &                  & $ 0.809\pm0.022$ &                  & \\
                &          CD &  1.93 &                  & $ 0.750\pm0.033$ &                  & \\
\tableline
Q2237$+$0305    &          AD &  1.01 & (variable;       & $ 0.130\pm0.145$ & $-0.008\pm0.068$ & \citet{wozniak}, \\
$\Rein =  0.85$ &          BD &  1.18 & see text)        & $ 0.172\pm0.139$ & $ 0.048\pm0.062$ & \citet{falco2237}, \\
cross           &          AC &  1.37 &                  & $ 0.289\pm0.170$ & $ 0.270\pm0.079$ & \citet{agol} \\
                &          BC &  1.40 &                  & $ 0.328\pm0.163$ & $ 0.319\pm0.072$ &
\enddata
\tablecomments{
The lengths $\Rein$ (Column 1) and $d_1$ (Column 3) are given in
arcseconds.  We do not explicitly quote $d_2$, because it can be
determined from the other $d_1$ values.  For example, in B0128+437
the value of $d_2$ for image pair AB would be the smaller of
$d_1(\mbox{AD})$ and $d_1(\mbox{CB})$.  CASTLES denotes the
CfA/Arizona Space Telescope Lens Survey
(see http://cfa-www.harvard.edu/castles).
}\label{tab:data}
\end{deluxetable}


\begin{thebibliography}{}

\bibitem[Agol et al.(2000)]{agol}
Agol, E., Jones, B., \& Blaes, O. 2000, \apj, 545, 657

\bibitem[Amara et al.(2004)]{amara}
Amara, A., Metcalf, R. B., Cox, T. J., \& Ostriker, J. P. 2004,
astro-ph/0411587

\bibitem[Bellman(1966)]{bellman}
Bellman, R. 1966, Perturbation Techniques in Mathematics, Engineering
and Physics (Mineola: Dover)

\bibitem[Bender et al.(1989)]{bender}
Bender, R., Surma, P., D\"obereiner, S., M\"ollenhoff, C., Madejski, R.
1989, \aap, 217, 35

\bibitem[Bernstein et al.(1997)]{0957bern}
Bernstein, G., Fischer, P., Tyson, J. A., \& Rhee, G. 1997, \apjl, 483, L79

\bibitem[Biggs et al.(2000)]{biggs}
Biggs, A. D., Xanthoupoulos, E., Browne, I. W. A., Koopmans, L. V. E.,
\& Fassnacht, C. D. 2000, \mnras, 318, 738

\bibitem[Blandford \& Narayan(1986)]{BN}
Blandford, R., \& Narayan, R. 1986, \apj, 310, 568

\bibitem[Brada\v{c} et al.(2002)]{bradac}
Brada\v{c}, M., Schneider, P., Steinmetz, M., Lombardi, M., King, L. J.,
\& Porcas, R. 2002, \aap, 388, 373

\bibitem[Brada\v{c} et al.(2004)]{bradac2}
Brada\v{c}, M., Schneider, P., Lombardi, M., Steinmetz, M., Koopmans, L. V. E.,
\& Navarro, J. F. 2004, \aap, 797, 809

\bibitem[Cardelli et al.(1989)Cardelli, Clayton \& Mathis]{cardelli}
Cardelli, J. A., Clayton, G. C., \& Mathis, J. S. 1989, \apj, 345, 245

\bibitem[Chen et al.(2003)Chen, Kravtsov \& Keeton]{chen}
Chen, J., Kravtsov, A. V., \& Keeton, C. R.  2003, \apj, 594, 24

\bibitem[Chiba(2002)]{chiba}
Chiba, M. 2002, \apj, 565, 17

\bibitem[Chiba et al.(2005)]{chibaIR}
Chiba, M., Minezaki, T., Kashikawa, N., Kataza, H., \& Inoue, K. T.
2005, \apj, 627, 53

\bibitem[Cohn et al.(2001)]{cohn}
Cohn, J. D., Kochanek, C. S., McLeod, B. A., \& Keeton, C. R. 2001, \apj,
554, 1216

\bibitem[Congdon \& Keeton(2005)]{congdon}
Congdon, A., \& Keeton, C. R. 2005, \apj, submitted

\bibitem[Dalal(1998)]{dalal}
Dalal, N. 1998, \apjl, 509, L13

\bibitem[Dalal \& Kochanek(2002)]{DK}
Dalal, N., \& Kochanek, C. S. 2002, \apj, 572, 25

\bibitem[Dalal \& Watson(2004)]{DW}
Dalal, N., \& Watson, C. R. 2004, astro-ph/0409483

\bibitem[Dobler \& Keeton(2005)]{dobler}
Dobler, G., \& Keeton, C. R. 2005, astro-ph/0502436

\bibitem[Evans \& Hunter(2002)]{evanshunter}
Evans, N. W., \& Hunter, C. 2002, \apj, 575, 68

\bibitem[Evans \& Witt(2003)]{EW}
Evans, N. W., \& Witt, H. J. 2003, \mnras, 345, 1351

\bibitem[Falco et al.(1996)]{falco2237}
Falco, E. E., Leh\'ar, J., Perley, R. A., Wambsganss, J., \& Gorenstein, M. V.
1996, \aj, 112, 897

\bibitem[Falco et al.(1999)]{falco}
Falco, E. E., et al. 1999, \apj, 523, 617

\bibitem[Fassnacht et al.(1999)]{fassnacht2045}
Fassnacht, C. D., et al. 1999, \aj, 117, 658

\bibitem[Fassnacht et al.(2002)]{fassnacht1608}
Fassnacht, C. D., Xanthopoulos, E., Koopmans, L. V. E., \& Rusin, D. 2002,
\apj, 581, 823

\bibitem[Fischer et al.(1998)]{fischer}
Fischer, P., Schade, D., \& Barrientos, L. P. 1998, \apj, 503, L127

\bibitem[Fluke \& Webster(1999)]{FW99}
Fluke, C. J., \& Webster, R. L. 1999, \mnras, 302, 68

\bibitem[Gaudi \& Petters(2002a)]{GPf}
Gaudi, B. S., \& Petters, A. O. 2002a, \apj, 574, 970

\bibitem[Gaudi \& Petters(2002b)]{GPc}
Gaudi, B. S., \& Petters, A. O. 2002b, \apj, 580, 468

\bibitem[Heyl et al.(1994)Heyl, Hernquist \& Spergel]{heyl}
Heyl, J. S., Hernquist, L., \& Spergel, D. N. 1994, \apj, 427, 165

\bibitem[Holder \& Schechter(2003)]{holder}
Holder, G., \& Schechter, P. 2003, \apj, 589, 688

\bibitem[Hunter \& Evans(2001)]{hunterevans}
Hunter, C., \& Evans, N. W. 2001, \apj, 554, 1227

\bibitem[Impey et al.(1998)]{impey1115}
Impey, C. D., et al. 1998, \apj, 509, 551

\bibitem[Inada et al.(2003)]{inada0924}
Inada, N., et al. 2003, \aj, 126, 666

\bibitem[Inada et al.(2004)]{inada1004}
Inada, N., et al. 2004, \pasj, submitted

\bibitem[J{\o}rgensen et al.(1995)J{\o}rgensen, Franx \& Kj{\ae}rgaard]{JFK}
J{\o}rgensen, I., Franx, M., \& Kj{\ae}rgaard, P. 1995, \mnras, 273, 1097

\bibitem[Keeton(2001)]{lenscode}
Keeton, C. R. 2001, astro-ph/0102340

\bibitem[Keeton(2003)]{analytics}
Keeton, C. R. 2003, \apj, 584, 664

\bibitem[Keeton et al.(2003)Keeton, Gaudi \& Petters]{cuspreln}
Keeton, C. R., Gaudi, B. S., \& Petters, A. O. 2003, \apj, 598, 138
(Paper I)

\bibitem[Keeton \& Kochanek(1997)]{kk1115}
Keeton, C. R., \& Kochanek, C. S. 1997, \apj, 487, 42

\bibitem[Keeton et al.(1997)Keeton, Kochanek \& Seljak]{KKS}
Keeton, C. R., Kochanek, C. S., \& Seljak, U. 1997, \apj, 482, 604

\bibitem[Keeton et al.(2000)]{0957host}
Keeton, C. R., et al. 2000, \apj, 542, 74 

\bibitem[Keeton \& Winn(2003)]{0134a}
Keeton, C. R., \& Winn, J. N. 2003, \apj, 590, 39

\bibitem[Keeton \& Zabludoff(2004)]{KZ04}
Keeton, C. R., \& Zabludoff, A. I. 2004, \apj, 612, 660

\bibitem[Keeton et al.(2005)]{hst0924}
Keeton, C. R., Burles, S., Schechter, P. L., \& Wambsganss, J. 2005,
astro-ph/0507521

\bibitem[Klypin et al.(1999)]{klypin}
Klypin, A., Kravtsov, A. V., Valenzuela, O., \& Prada, F. 1999, \apj, 522, 82

\bibitem[Kochanek(1991)]{csk91}
Kochanek, C. S. 1991, \apj, 373, 354

\bibitem[Kochanek(2004a)]{csk2237}
Kochanek, C. S. 2004a, 605, 58

\bibitem[Kochanek(2004b)]{csk0924}
Kochanek, C. S. 2004b, in The Impact of Gravitational Lensing on
Cosmology (IAU 225), eds. Y. Mellier \& G. Meylan; also astro-ph/0412089

\bibitem[Kochanek \& Dalal(2004)]{KD}
Kochanek, C. S., \& Dalal, N. 2004, \apj, 610, 69


\bibitem[Koopmans et al.(2003a)]{koopmans}
Koopmans, L. V. E., et al. 2003a, \apj, 595, 712

\bibitem[Koopmans et al.(2003b)]{koopmans1608}
Koopmans, L. V. E., et al. 2003b, \apj, 599, 70

\bibitem[Marlow et al.(1999)]{marlow}
Marlow, D. R., et al. 1999, \aj, 118, 654

\bibitem[Mao(1992)]{mao92}
Mao, S. 1992, \mnras, 389, 63

\bibitem[Mao \& Schneider(1998)]{MS}
Mao, S., \& Schneider, P. 1998, \mnras, 295, 587

\bibitem[Mao et al.(2004)]{MaoJing}
Mao, S., Jing, Y., Ostriker, J. P., \& Weller, J. 2004, \apj, 604, L5

\bibitem[Metcalf(2005a)]{metcalfLOS1}
Metcalf, R. B. 2005a, \apj, 622, 72

\bibitem[Metcalf(2005b)]{metcalfLOS2}
Metcalf, R. B. 2005b, \apj, in press (astro-ph/0412538)

\bibitem[Metcalf \& Madau(2001)]{MM}
Metcalf, R. B., \& Madau, P.. 2001, \apj, 563, 9

\bibitem[Metcalf \& Zhao(2002)]{MZ}
Metcalf, R. B., \& Zhao, H. 2002, \apj, 567, L5

\bibitem[Metcalf et al.(2004)]{metcalf2237}
Metcalf, R. B., Moustakas, L. A., Bunker, A. J., \& Parry, I. R.
2004, \apj, 607, 43

\bibitem[M{\"o}ller et al.(2003)]{moeller}
M{\"o}ller, O., Hewett, P., \& Blain, A. W. 2003, \mnras, 345, 1

\bibitem[Moore et al.(1999)]{bmoore}
Moore, B., Ghigna, S., Governato, F., Lake, G., Quinn, T., Stadel, J.,
\& Tozzi, P. 1999, \apj, 524, L19

\bibitem[Moore \& Hewitt(1997)]{moore}
Moore, C. B., \& Hewitt, J. N. 1997, \apj, 491, 451

\bibitem[Morgan et al.(2004)]{morgan}
Morgan, N. D., et al. 2004a, \aj, 127, 2617

\bibitem[Morgan et al.(2005)]{morgan0435}
Morgan, N. D., Kochanek, C. S., Pevunova, O., \& Schechter, P. L.
2005, \aj, 129, 2531

\bibitem[Moustakas \& Metcalf(2003)]{moustakas}
Moustakas, L. A., \& Metcalf, R. B. 2003, \mnras, 339, 607

\bibitem[Naab \& Burkert(2003)]{naab}
Naab, T., \& Burkert, A. 2003, \apj, 597, 893

\bibitem[Oguri \& Lee(2004)]{oguri}
Oguri, M., \& Lee, J. 2004, \mnras, 355, 120

\bibitem[Oguri et al.(2004)]{oguri1004}
Oguri, M., et al. 2004, \apj, 605, 78

\bibitem[Patnaik et al.(1999)]{patnaik}
Patnaik, A. R., Kemball, A. J., Porcas, R. W., \& Garrett, M. A. 1999,
\mnras, 307, L1

\bibitem[Patnaik \& Narasimha(2001)]{patnaik2}
Patnaik, A. R., \& Narasimha, D. 2001, \mnras, 326, 1403 

\bibitem[Petters et al.(2001)Petters, Levine \& Wambsganss]{PLW}
Petters, A. O., Levine, H., \& Wambsganss, J. 2001, Singularity Theory
and Gravitational Lensing (Boston: Birkh\"auser)

\bibitem[Phillips et al.(2000)]{phillips}
Phillips, P. M., et al. 2000, \mnras, 319, L7

\bibitem[Quadri et al.(2003)]{quadri}
Quadri, R., M{\"o}ller, O., \& Natarajan, P. 2003, \apj, 597, 659

\bibitem[Ratnatunga et al.(1995)]{ratnatunga}
Ratnatunga, K., Ostrander, E. J., Griffiths, R. E., \& Im, M. 1995,
\apj, 453, L5

\bibitem[Rest et al.(2001)]{rest}
Rest, A., van den Bosch, F. C., Jaffe, W., Tran, H., Tsvetanov, Z.,
Ford, H. C., Davies, J., \& Schafer, J. 2001, \aj, 121, 2431

\bibitem[Richards et al.(2004)]{richards}
Richards, G. T., et al. 2004, \apj, 610, 679

\bibitem[Rusin \& Tegmark(2001)]{RT}
Rusin, D., \& Tegmark, M. 2001, \apj, 553, 709

\bibitem[Rusin et al.(2001)]{1359}
Rusin, D., et al. 2001, \apj, 557, 594

\bibitem[Saglia et al.(1993)Saglia, Bender \& Dressler]{saglia}
Saglia, R. P., Bender, R., \& Dressler, A. 1993, \aap, 279, 75

\bibitem[Saha \& Williams(2003)]{saha}
Saha, P., \& Williams, L. L. R. 2003, \aj, 125, 2769

\bibitem[Schechter \& Moore(1993)]{SchechterMoore}
Schechter, P. L., \& Moore, C. B. 1993, \aj, 105, 1

\bibitem[Schechter \& Wambsganss(2002)]{SW}
Schechter, P. L., \& Wambsganss, J. 2002, \apj, 580, 685

\bibitem[Schechter et al.(2003)]{s1104}
Schechter, P. L., Udalski, A., Szyma{\'n}ski, M., Kubiak, M.,
Pietrzy{\'n}ski, G., Soszy{\'n}ski, I., Wo{\'z}niak, P., {\. Z}ebru{\'n}, K.,
Szewczyk, O., \& Wyrzykowski, {\L}. 2003, \apj, 584, 657

\bibitem[Schneider et al.(1992)Schneider, Ehlers \& Falco]{SEF}
Schneider, P., Ehlers, J., \& Falco, E. E. 1992, Gravitational Lenses
(Berlin: Springer) 

\bibitem[Schneider \& Weiss(1992)]{weiss}
Schneider, P., \& Weiss, A. 1992, \aap, 260, 1

\bibitem[Sluse et al.(2003)]{sluse}
Sluse, D., et al. 2003, \aap, 406, L43

\bibitem[Sykes et al.(1998)]{sykes}
Sykes, C. M., et al. 1998, \mnras, 301, 310

\bibitem[Trotter et al.(2000)Trotter, Winn \& Hewitt]{trotter}
Trotter, C. S., Winn, J. N., \& Hewitt, J. N. 2000, \apj, 535, 671

\bibitem[Warren et al.(1996)]{warren96}
Warren, S. J., Hewett, P. C., Lewis, G. F., M\o ller, P., Iovino, A.,
\& Shaver, P. A. 1996, \mnras, 278, 139

\bibitem[Warren et al.(1999)]{warren99}
Warren, S. J., Lewis, G. F., Hewett, P. C., M\o ller, P., Shaver, P. A., \&
Iovino, A.  1999, \aap, 343, L35

\bibitem[Wayth et al.(2005a)]{wayth}
Wayth, R. B., O'Dowd, M., \& Webster, R. L. 2005, \mnras, 359, 561

\bibitem[Wayth et al.(2005b)]{wayth0047}
Wayth, R. B., Warren, S. J., Lewis, G. F., \& Hewett, P. C. 2005b,
\mnras, 360, 1333

\bibitem[Winn et al.(2003)]{0134b}
Winn, J. N., Kochanek, C. S., Keeton, C. R., \& Lovell, J. E. J.
2003, \apj, 590, 26

\bibitem[Wisotzki et al.(1999)]{wisotzki0230}
Wisotzki, L., Christleib, N., Liu, M. C., Maza, J., Morgan, N. D., \&
Schechter, P. L. 1999, \aap, 348, L41

\bibitem[Wisotzki et al.(2002)]{wisotzki0435-1}
Wisotzki, L., Schechter, P. L., Bradt, H. V., Heinm\"uller, J., \& Reimers, D.
2002, \aap, 395, 17

\bibitem[Wisotzki et al.(2003)]{wisotzki0435-2}
Wisotzki, L., Becker, T., Christensen, L., Helms, A. Jahnke, K., Kelz, A.,
Roth, M. M., \& Sanchez, S. F. 2003, \aap, 408, 455

\bibitem[Witt \& Mao(1997)]{wittmao}
Witt, H. J., \& Mao, S. 1997, \mnras, 291, 211

\bibitem[Witt \& Mao(2000)]{wittmao00}
Witt, H. J., \& Mao, S. 2000, \mnras, 311, 689

\bibitem[Wo\'zniak et al.(2000)]{wozniak}
Wo\'zniak, P. R., Alard, C. Udalski, A., Szyma\'nski, M., Kubiak, M.,
Pietrzy\'nski, G., \& Zebru\'n, K. 2000, \apj, 529, 88

\bibitem[Yoo et al.(2005)]{yoo}
Yoo, J., Kochanek, C. S., Falco, E. E., \& McLeod, B. A. 2005,
\apj, 626, 51

\bibitem[Zentner \& Bullock(2003)]{zentner}
Zentner, A. R., \& Bullock, J. S. 2003, \apj, 598, 49

\end{thebibliography}
\end{document}